\documentclass[]{aa}
\usepackage{epsfig}

\def\gsim{\vcenter{\hbox{$>$}\offinterlineskip\hbox{$\sim$}}}
\begin{document}
\title{Very Large Telescope three micron spectra of dust-enshrouded red giants
       in the Large Magellanic Cloud}
%\footnote{Based on observations collected at the European Southern
% Observatory, Chile (ESO N$^{\rm o}$ 72.D-0351).}
\author{Jacco Th. van Loon\inst{1}, Jonathan R. Marshall\inst{1}, Martin
        Cohen\inst{2}, Mikako Matsuura\inst{3}\\ Peter R. Wood\inst{4},
        Issei Yamamura\inst{5} and Albert A. Zijlstra\inst{3}}
\institute{Astrophysics Group, School of Physical \& Geographical Sciences,
           Keele University, Staffordshire ST5 5BG, UK
      \and Radio Astronomy Lab, 601 Campbell Hall, University of California at
           Berkeley, Berkeley CA 94720-3411, USA
      \and Department of Physics and Astronomy, University of Manchester,
           Sackville Street, P.O.Box 88, Manchester M60 1QD, UK
      \and Research School of Astronomy and Astrophysics, Australian National
           University, Cotter Road, Weston Creek, ACT 2611, Australia
      \and Institute of Space and Astronautical Science, 3-1-1 Yoshinodai,
           Sagamihara, Kanagawa 229, Japan
}
%\offprints{\email{jacco@astro.keele.ac.uk}}
\date{Received date; accepted date}
\titlerunning{3 $\mu$m spectra of dust-enshrouded stars in the LMC}
\authorrunning{van Loon et al.}
\abstract{We present ESO/VLT spectra in the 2.9--4.1 $\mu$m range for a large
sample of infrared stars in the Large Magellanic Cloud (LMC), selected on the
basis of MSX and 2MASS colours to be extremely dust-enshrouded AGB star
candidates. Out of 30 targets, 28 are positively identified as carbon stars,
significantly adding to the known population of optically invisible carbon
stars in the LMC. We also present spectra for six IR-bright stars in or near
three clusters in the LMC, identifying four of them as carbon stars and two as
oxygen-rich supergiants. We analyse the molecular bands of C$_2$H$_2$ at 3.1
and 3.8 $\mu$m, HCN at 3.57 $\mu$m, and sharp absorption features in the
3.70--3.78 $\mu$m region that we attribute to C$_2$H$_2$. There is evidence
for a generally high abundance of C$_2$H$_2$ in LMC carbon stars, suggestive
of high carbon-to-oxygen abundance ratios at the low metallicity in the LMC.
The low initial metallicity is also likely to have resulted in less abundant
HCN and CS. The sample of IR carbon stars exhibits a range in C$_2$H$_2$:HCN
abundance ratio. We do not find strong correlations between the properties of
the molecular atmosphere and circumstellar dust envelope, but the observed
differences in the strengths and shapes of the absorption bands can be
explained by differences in excitation temperature. High mass-loss rates and
strong pulsation would then be seen to be associated with a large scale height
of the molecular atmosphere.
\keywords{
Stars: carbon --
Stars: AGB and post-AGB --
circumstellar matter --
Stars: mass-loss --
Magellanic Clouds --
Infrared: stars}}
\maketitle

%=========================================================================== 1
\section{Introduction}

One of the main contributors of elements such as carbon and nitrogen, and
arguably the most important ``factory'' of cosmic dust, Asymptotic Giant
Branch (AGB) stars represent the final evolutionary stages of
intermediate-mass stars ($M_{\rm ZAMS}\la1$ to $\sim8$ M$_\odot$), when they
lose up to 80 per cent of their mass at rates of $\dot{M}\sim10^{-6}$ to
$10^{-4}$ M$_\odot$ yr$^{-1}$. Aged between $t\sim30$ Myr and $t\ga10$ Gyr,
AGB stars chemically enrich the Universe on timescales ranging from less than
the dynamical timescale of a galaxy to as long as the age of the Universe. AGB
mass loss happens as a result of two mechanisms: (1) strong radial pulsations
elevate the stellar atmosphere out to a distance where (2) the temperature is
sufficiently low ($T\sim1000$ K) and the density is sufficiently high
($n_{{\rm H}_2}\gsim10^{13}$ m$^{-3}$) for dust formation to occur. Radiation
pressure from the luminous giant ($L\sim10^4$ L$_\odot$) drives away the dust,
taking the gas along with it by mutual collisions. During this ``superwind''
stage the dust-enshrouded AGB star vanishes at optical wavelengths but shines
brightly in the infrared.

An unsolved problem of AGB mass loss is how the simple molecules in the stellar
photosphere grow into larger assemblies that form the cores for dust
condensation, and how efficient these processes are. Little is known about the
atmospheric chemistry, the molecular abundances within the dust-formation
zone, and the dust-to-gas ratio in the wind. How do molecular chemistry and
dust condensation depend on the stellar mass, luminosity, temperature,
pulsation period and amplitude, on the initial metallicity and on the
self-enrichment of the photosphere as a result of the dredge-up of
nucleosynthesis products? We do know that the transformation of some
oxygen-rich AGB stars into carbon stars (by dredge-up on the upper AGB)
changes the molecular and dust formation chemistry from oxygen-dominated to
carbon-dominated, giving rise to a vastly different array of molecules and
dust particles. Hence these two types of AGB star enrich the interstellar
medium with completely different material.

The Large and Small Magellanic Clouds (LMC and SMC) offer an excellent
opportunity to study AGB mass loss, at metallicities of $\sim40$ and 15 per
cent solar, respectively. Accurate luminosities and mass-loss rates can be
obtained in these galaxies as a result of their proximity and known distances
of 50 and 60 kpc, respectively. Much work on magellanic dust-enshrouded AGB
stars has been done (e.g., Wood et al.\ 1992; Zijlstra et al.\ 1996; van Loon
et al.\ 1997, 1998). The mass-loss rate depends strongly on luminosity but
also on temperature and pulsation properties (Wood 1998; van Loon et al.\
1999, 2005; Marshall et al.\ 2004; Whitelock et al.\ 2003). Dust mass-loss
rates are lower in metal-poor environments, but this may not be true for the
gas mass-loss rates (van Loon 2000).

Metal-poor carbon stars are surprisingly abundant in C$_2$ and C$_2$H$_2$, and
may have a higher C/O ratio than their galactic solar-metallicity equivalents
(van Loon, Zijlstra \& Groenewegen 1999; Matsuura et al.\ 2002, 2005). This
may be due to the high contrast between the carbon self-enrichment and
oxygen-poor photosphere of metal-poor carbon stars, and could have important
consequences for the condensation of carbonaceous dust grains in
low-metallicity environments.

To investigate the formation of simple, di-atomic molecules such as C$_2$ and
CN, the assembly into slightly more complex molecules such as C$_2$H$_2$ and
HCN, and the condensation into dust grains, we set out to measure the
molecular abundances in the extended atmospheres of mass-losing carbon-rich
AGB stars in the LMC. Because many of these highly-evolved objects are
dust-enshrouded, it is impossible to perform such a study at optical
wavelengths. Furthermore, whilst the optical spectrum is rich in bands of
di-atomic molecules, larger molecules are observed at infrared wavelengths.
The three micron telluric window (2.9--4.1 $\mu$m) in carbon stars shows
distinctive broad and strong absorption by C$_2$H$_2$ and HCN molecules at 3.1
$\mu$m, and often absorption by C$_2$H$_2$ at 3.8 $\mu$m and HCN at 3.57
$\mu$m. These molecules are formed in the upper atmospheric layers by
hydration of the di-atomic molecules that are seen at optical wavelengths.
C$_2$H$_2$ is a possible building block for Polycyclic Aromatic Hydrocarbons
(PAHs) (Latter 1991), and must play a key r\^{o}le in the condensation of
carbonaceous grains in H-rich environments (Gail \& Sedlmayr 1988). The HCN
band strength may be a measure of the nitrogen abundance.

Here we present ESO/VLT three micron spectra of a large sample of heavily
dust-enshrouded carbon stars in the LMC, and an analysis of their molecular
atmospheres.

%=========================================================================== 2
\section{Observations}

\subsection{Targets}

The targets for our 3--4 $\mu$m spectroscopy were drawn from a sample of
mid-IR sources detected with the Mid-course Space eXperiment (MSX) (Egan, Van
Dyk \& Price 2001; Wood \& Cohen 2001). Our principal objective was to
establish the nature and chemical composition of heavily dust-enshrouded stars
for which optical spectroscopy would not be possible, to help select targets
for our Spitzer Space Telescope programme \#3505 (PI: P.R.\ Wood). Hence we
selected objects by excluding all objects that have optical photometry in the
MACHO database, bright objects with a K$_{\rm s}$-band magnitude $<9$ mag, and
objects that are not extremely red with a J--K$_{\rm s}$ colour $<5$ mag. The
actually observed objects are mostly of $8^{\rm th}$ or $9^{\rm th}$ magnitude
in the 3--4 $\mu$m range. Throughout the paper we refer to the MSX-selected
sample by their MSX\,LMC catalogue numbers.

Six more dusty objects were observed that were known to be in or near the
populous LMC clusters NGC\,1903, NGC\,1984 and NGC\,1978 (van Loon, Marshall
\& Zijlstra 2005). VLT/ISAAC spectra covering the same spectral region are
available from the literature for nine more IR objects in the LMC (Matsuura et
al.\ 2005). Of these, IRAS\,04496$-$6958, IRAS\,04557$-$6753 and LI-LMC\,1813
are cluster members (van Loon et al.\ 2005b). The IR photometric data of the
cluster objects can be found in van Loon et al.\ (2005b) and are not repeated
here.

\subsection{Spectroscopy}

The ISAAC instrument on the European Southern Observatory (ESO) Very Large
Telescope (VLT) ``Antu'' at Paranal, Chile, was used on the nights of 6, 7 and
8 December 2003 to obtain low-resolution long-slit spectra between 2.9 and 4.1
$\mu$m. At a slit width of $2^{\prime\prime}$ the spectral resolving power was
essentially determined by the seeing and was $300{\la}R\la700$. The smallest
attained resolution element is sampled by three pixels on the Aladdin array.

The spectra were obtained by chopping and nodding with a throw of
$10^{\prime\prime}$ ($15^{\prime\prime}$ in some cases to avoid chopping onto
another star) to cancel the high background at these thermal-IR wavelengths,
and jittering within $2^{\prime\prime}$ to remove the effects of bad pixels.
This produced three spectra on the final combined frame, with the central
spectrum twice as bright as the other (inverse) spectra. Exposure times varied
between 12 and 60 minutes. After removing cosmic ray hits and correcting the
spectra for their small inclination with respect to the array, all spectra
were extracted using an optimal extraction algorithm. The extracted spectra
were then aligned and combined. Spectra of an internal Xe+Ar lamp were used to
calibrate the wavelengths; the spectra are sampled on a grid of 14 \AA\
elements.

%
% TABLE 1
%
\begin{table*}
\caption[]{List of targets, in order of increasing Right Ascension (2MASS
coordinates, in J2000). MSX\,LMC (Egan, van Dyk \& Price 2001; targets for
Spitzer Space Telescope programmes are in boldface) and IRAS\,PSC designations
are given where available, together with near-IR magnitudes (1=2MASS, 2=MSSSO,
3=spectroscopy acquisition images) and MSX band-A and IRAS 12 and 25 $\mu$m
flux densities (in mJy). Values marked with a colon are suspect. References to
the identification as a dust-enshrouded AGB star are as follows:
1=Egan, Van Dyk \& Price (2001);
2=Loup et al.\ (1997);
3=Reid, Tinney \& Mould (1990);
4=Trams et al.\ (1999);
5=van Loon et al.\ (1997);
6=van Loon (2000);
7=Zijlstra et al.\ (1996);
8=this work.}
\begin{tabular}{rcrrrrrrrrrrrrl}
\hline\hline
MSX                                             &
IRAS                                            &
$\alpha$ ($^{\rm h}$ $^{\rm m}$ $^{\rm s}$)     &
$\delta$ ($^\circ$ $^\prime$ $^{\prime\prime}$) &
J(1)                                            &
H(1)                                            &
K$_{\rm s}$(1\rlap{)}                           &
J(2)                                            &
K(2)                                            &
L(2)                                            &
\llap{L}$^\prime$(3)                            &
F$_{\rm A}$                                     &
F$_{12}$                                        &
F$_{25}$                                        &
Re\rlap{f}                                      \\
\hline
\multicolumn{14}{l}{\it IR carbon stars} \\
1042                  &
04374$-$6831          &
4 37 22.7             &
\llap{$-$}68 25 03    &
16.33                 &
13.76                 &
\llap{1}1.55          &
                      &
                      &
                      &
 9.31                 &
107                   &
240                   &
120                   &
2,7                   \\
1249                  &
04552$-$6536          &
4 55 27.6             &
\llap{$-$}65 31 07    &
14.15                 &
12.27                 &
\llap{1}0.77          &
                      &
                      &
                      &
 8.18                 &
216                   &
160                   &
110                   &
2,5                   \\
  50                  &
05026$-$6809          &
5 02 22.0             &
\llap{$-$}68 05 24    &
$>$17.9               &
$>$16.1               &
\llap{1}4.67          &
                      &
\llap{1}2.44          &
                      &
 8.78                 &
194                   &
250\rlap{:}           &
200\rlap{:}           &
2,5                   \\
  91                  &
05038$-$6857          &
5 03 38.5             &
\llap{$-$}68 53 13    &
$>$16.4               &
15.56                 &
\llap{1}2.87          &
\llap{1}6.67          &
\llap{1}1.25          &
                      &
 9.24                 &
238                   &
420                   &
250                   &
1                     \\
  93                  &
                      &
5 05 27.9             &
\llap{$-$}68 56 36    &
16.83                 &
14.36                 &
\llap{1}2.01          &
                      &
\llap{1}3.01          &
 9.92                 &
 8.82                 &
170                   &
160                   &
100                   &
1                     \\
  55                  &
05058$-$6843          &
5 05 41.7             &
\llap{$-$}68 39 11    &
                      &
                      &
                      &
                      &
\llap{1}1.24          &
                      &
 9.25                 &
161                   &
260                   &
170                   &
8                     \\
  45                  &
05108$-$6839          &
5 10 41.3             &
\llap{$-$}68 36 07    &
$>$16.3               &
14.56                 &
\llap{1}1.74          &
                      &
\llap{1}2.82          &
                      &
 7.87                 &
372                   &
430                   &
380                   &
1                     \\
{\bf 47}\rlap{$^d$}   &
05113$-$6739          &
5 11 13.9             &
\llap{$-$}67 36 16    &
$>$17.7               &
14.77                 &
\llap{1}2.49          &
                      &
                      &
                      &
 8.02                 &
233                   &
250                   &
140                   &
7                     \\
{\bf 219}\rlap{$^d$}  &
                      &
5 11 19.5             &
\llap{$-$}68 42 28    &
$>$15.6               &
$>$14.7               &
\llap{1}3.14          &
\llap{1}7.93          &
\llap{1}2.31          &
                      &
 8.62                 &
204                   &
180                   &
120                   &
1                     \\
 232                  &
                      &
5 12 04.0             &
\llap{$-$}69 16 21    &
$>$16.9               &
14.93                 &
\llap{1}2.84          &
\llap{1}7.78          &
\llap{1}2.52          &
 9.73                 &
 9.23                 &
135                   &
220                   &
120                   &
1                     \\
{\bf 223}\rlap{$^d$}  &
05132$-$6941          &
5 12 51.1             &
\llap{$-$}69 37 50    &
$>$17.8               &
$>$15.7               &
\llap{1}3.36          &
                      &
\llap{1}3.00          &
 9.34                 &
 8.33                 &
259                   &
220                   &
130                   &
1                     \\
 202                  &
                      &
5 14 14.9             &
\llap{$-$}70 04 10    &
$>$18.1               &
15.92                 &
\llap{1}3.07          &
\llap{1}7.81          &
\llap{1}2.36          &
                      &
 9.76                 &
176                   &
120                   &
70                    &
1                     \\
 225                  &
                      &
5 14 37.9             &
\llap{$-$}68 19 21    &
$>$18.0               &
$>$16.7               &
\llap{1}3.42          &
                      &
\llap{1}2.73          &
                      &
 8.48                 &
215                   &
200                   &
140                   &
1                     \\
 221                  &
                      &
5 15 15.7             &
\llap{$-$}69 00 34    &
$>$16.1               &
14.92                 &
\llap{1}2.74          &
\llap{1}7.36          &
\llap{1}1.93          &
 9.37                 &
 9.10                 &
189                   &
260                   &
220                   &
1                     \\
{\bf 349}\rlap{$^d$}  &
                      &
5 17 26.9             &
\llap{$-$}68 54 58    &
$>$17.4               &
$>$16.9               &
\llap{1}4.82          &
                      &
\llap{1}3.49          &
 9.81                 &
 9.04                 &
193                   &
220                   &
130                   &
1                     \\
{\bf 307}\rlap{$^d$}  &
05190$-$6748          &
5 18 56.3             &
\llap{$-$}67 45 04    &
$>$18.2               &
$>$15.9               &
\llap{1}3.10          &
                      &
                      &
                      &
 8.62                 &
209                   &
390                   &
250                   &
3                     \\
{\bf 341}\rlap{$^d$}  &
                      &
5 21 00.4             &
\llap{$-$}69 20 55    &
$>$17.5               &
15.92                 &
\llap{1}3.15          &
\llap{1}7.47          &
\llap{1}2.08          &
 9.44                 &
 9.59                 &
210                   &
100                   &
60                    &
1                     \\
 484                  &
                      &
5 25 03.3             &
\llap{$-$}69 26 17    &
$>$15.2               &
$>$14.6               &
\llap{1}4.48          &
                      &
\llap{1}3.77          &
\llap{1}0.59          &
 9.43                 &
156                   &
\llap{$<$}100         &
\llap{$<$}100         &
1                     \\
 774                  &
                      &
5 26 23.1             &
\llap{$-$}69 11 20    &
$>$16.8               &
15.08                 &
\llap{1}2.72          &
\llap{1}8.00          &
\llap{1}2.49          &
 9.54                 &
 8.87                 &
237                   &
\llap{$<$}250         &
140\rlap{:}           &
1                     \\
 677                  &
                      &
5 27 11.0             &
\llap{$-$}69 34 14    &
$>$16.2               &
14.01                 &
\llap{1}1.90          &
\llap{1}6.99          &
\llap{1}1.69          &
 9.18                 &
 8.82                 &
105                   &
180                   &
90                    &
1                     \\
{\bf 635}\rlap{$^d$}  &
05278$-$6942          &
5 27 24.1             &
\llap{$-$}69 39 45    &
$>$17.5               &
$>$15.6               &
\llap{1}2.35          &
\llap{1}8.13          &
\llap{1}1.98          &
 8.45                 &
 9.60                 &
684                   &
450                   &
380                   &
2,5                   \\
{\bf 692}\rlap{$^d$}  &
05295$-$7121          &
5 28 46.6             &
\llap{$-$}71 19 13    &
16.80                 &
14.32                 &
\llap{1}2.18          &
                      &
                      &
                      &
 8.67                 &
204                   &
230                   &
80                    &
2,7                   \\
 644                  &
                      &
5 29 22.9             &
\llap{$-$}70 06 46    &
$>$16.4               &
14.28                 &
\llap{1}1.93          &
\llap{1}7.46          &
\llap{1}1.92          &
 8.72                 &
 8.34                 &
232                   &
160                   &
100                   &
1                     \\
 654                  &
                      &
5 30 05.1             &
\llap{$-$}69 56 46    &
$>$14.4               &
$>$13.8               &
\llap{1}3.10          &
                      &
\llap{1}2.39          &
 9.36                 &
 9.81                 &
152                   &
200\rlap{:}           &
100\rlap{:}           &
1                     \\
1780                  &
                      &
5 33 01.8             &
\llap{$-$}68 23 59    &
$>$17.9               &
$>$17.0               &
\llap{1}4.10          &
                      &
\llap{1}3.76          &
                      &
 8.74                 &
309                   &
400                   &
\llap{$<$}600         &
1                     \\
{\bf 743}\rlap{$^d$}  &
                      &
5 34 53.7             &
\llap{$-$}70 29 25    &
$>$14.1               &
$>$13.2               &
\llap{1}3.27          &
\llap{1}8.65          &
\llap{1}2.82          &
 9.67                 &
 9.37                 &
165                   &
260                   &
100\rlap{:}           &
1                     \\
{\bf 872}\rlap{$^d$}  &
05360$-$6648          &
5 36 01.2             &
\llap{$-$}66 46 40    &
$>$18.0               &
16.07                 &
\llap{1}3.28          &
                      &
                      &
                      &
 8.64                 &
192                   &
210                   &
90                    &
3                     \\
 737                  &
                      &
5 36 36.8             &
\llap{$-$}70 43 46    &
16.58                 &
13.72                 &
\llap{1}1.42          &
\llap{1}6.32          &
\llap{1}1.27          &
                      &
 8.31                 &
153                   &
310                   &
150\rlap{:}           &
1                     \\
\hline
\multicolumn{14}{l}{\it IR objects of unknown type} \\
{\bf 196}\rlap{$^b$}  &
05125$-$7035          &
5 12 00.8             &
\llap{$-$}70 32 24    &
$>$18.0               &
$>$16.8               &
\llap{1}4.52          &
                      &
                      &
                      &
\llap{1}0.59          &
451                   &
500                   &
500                   &
2,5                   \\
{\bf 733}\rlap{$^a$}  &
05348$-$7024          &
5 34 16.0             &
\llap{$-$}70 22 53    &
$>$17.7               &
15.49                 &
\llap{1}2.85          &
                      &
\llap{1}2.70          &
 9.28                 &
 9.44                 &
414                   &
580                   &
160                   &
2,7                   \\
\hline
\multicolumn{14}{l}{\it LMC carbon-rich stars from Matsuura et al.\ (2005)} \\
{\bf 1007}\rlap{$^c$} &
04286$-$6937          &
4 28 30.2             &
\llap{$-$}69 30 50    &
\llap{16.35}          &
\llap{13.87}          &
\llap{11.86}          &
                      &
                      &
8.84\rlap{$^A$}       &
                      &
155                   &
280                   &
200                   &
2,7                   \\
1198                  &
04539$-$6821          &
4 53 46.4             &
\llap{$-$}68 16 13    &
$>$18.1               &
\llap{15.44}          &
\llap{13.03}          &
                      &
                      &
9.50\rlap{$^A$}       &
                      &
193                   &
220                   &
120                   &
2,7                   \\
{\bf 44}\rlap{$^d$}   &
05112$-$6755          &
5 11 10.5             &
\llap{$-$}67 52 11    &
\llap{16.41}          &
\llap{14.07}          &
\llap{11.69}          &
                      &
                      &
9.13\rlap{$^A$}       &
                      &
451                   &
460                   &
330                   &
3                     \\
--\rlap{$^B$}         &
                      &
5 20 46.7             &
\llap{$-$}69 01 24    &
\llap{11.43}          &
\llap{10.45}          &
9.77                  &
                      &
                      &
8.80\rlap{$^A$}       &
                      &
65                    &
60\rlap{:}            &
20\rlap{:}            &
4                     \\
\hline
\multicolumn{14}{l}{\it LMC oxygen-rich stars from Matsuura et al.\ (2005)} \\
{\bf 283}\rlap{$^a$}  &
05128$-$6455          &
5 13 04.6             &
\llap{$-$}64 51 40    &
\llap{14.55}          &
\llap{12.83}          &
\llap{11.28}          &
                      &
                      &
9.00\rlap{$^A$}       &
                      &
259                   &
230                   &
240                   &
2,7                   \\
{\bf 264}\rlap{$^c$}  &
05148$-$6730\rlap{$^C$} &
5 14 49.7             &
\llap{$-$}67 27 20    &
8.64                  &
7.78                  &
7.42                  &
                      &
                      &
6.97\rlap{$^A$}       &
                      &
331                   &
440                   &
230                   &
3                     \\
\hline
\end{tabular}\\
Spitzer programmes: $a=$ GTO \#200 (Houck); $b=$ DDT \#1094 (Kemper); $c=$ GO
\#3426 (Kastner); $d=$ GO \#3505 (Wood).\\
Notes: $A=$ SAAO monitoring at 3.5 $\mu$m (Whitelock et al.\ 2003); $B=$
BMB-R\,46, SHV\,0521050$-$690415; $C=$ HV\,916.
\end{table*}

The relative spectral response calibration was obtained by dividing by the
spectrum of the B-type standard star HIP\,020020. This removed most of the
telluric absorption lines --- except the saturated methane line at 3.32 $\mu$m
--- but introduced artificial emission features due to photospheric absorption
lines in the spectrum of HIP\,020020. Each spectrum was therefore multiplied
by a synthetic spectrum appropriate for HIP\,020020, in which we have inserted
absorption lines due to Br$\alpha$ 4.052, Pf$\gamma$ 3.741, Pf$\delta$ 3.297,
Pf$\epsilon$ 3.039 and Pf$\zeta$ 2.873 $\mu$m of depths sufficient to remove
the spurious emission lines. Small wavelength shifts and differences in
effective resolution can cause imperfections in the correction. Fig.\ 1 shows
the observed uncorrected spectra of HIP\,020020 and target star MSX\,LMC\,47,
a low-resolution Kurucz spectrum ($T_{\rm eff}=18,000$ K, $\log{g}=4.0$ and
[Fe/H]=0) and our synthetic spectrum. The observed spectra show the extent to
which the telluric absorption affects the spectrum especially around 3.2--3.3
$\mu$m, and the effective band width between 2.85 and 4.15 $\mu$m.

%
% FIGURE 1
%
\begin{figure}[tb]
\centerline{\psfig{figure=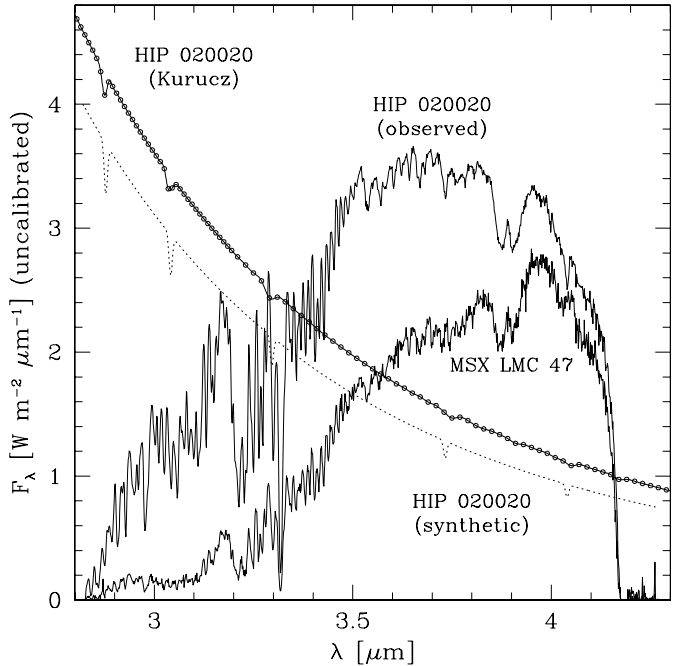,width=88mm}}
\caption[]{Uncorrected ESO-VLT/ISAAC spectra of the B-type standard star
HIP\,020020 and a target star, MSX\,LMC\,47, plus a low-resolution Kurucz
spectrum and our synthetic spectrum appropriate for the photosphere of
HIP\,020020.}
\end{figure}

A $\kappa$-$\sigma$ clipping algorithm was applied to remove spikes from the
spectrum that are caused by hot pixels. Care was taken to ensure that
astrophysical features were not affected by this procedure.

The conditions were very similar for all three nights. The seeing as measured
on the acquisition images was around $0.5^{\prime\prime}$ most of the time but
occasionally worsened up to about $1^{\prime\prime}$ (e.g., when observing at
high airmass of $\sim$2). The relative humidity and air temperature were
fairly constant at 8--18 per cent and 12--15$^\circ$C, respectively.

%
% TABLE 2
%
\begin{table*}
\caption[]{Additional data from DENIS and ISO, and pulsation properties from
near-IR monitoring. The flux densities are in mJy, pulsation periods in days
and K$_{\rm s}$-band amplitudes in magnitudes. The references are as follows:
1=Loup et al.\ (in preparation);
2=Trams et al.\ (1999);
3=van Loon et al.\ (2005a);
4=van Loon et al.\ (2005b);
5=Whitelock et al.\ (2003);
6=Wood (1998).}
\begin{tabular}{lrrrrrrrrlrrl}
\hline\hline
Name                 &
I$_{\rm DENIS}$      &
J$_{\rm DENIS}$      &
K$_{\rm s, DENIS}$   &
F$_{4.5}$            &
F$_{6.7}$            &
F$_{12}$             &
F$_{14.3}$           &
F$_{25}$             &
Spectrum             &
P(d)                 &
$\Delta$K$_{\rm s}$  &
Ref                  \\
\hline
\multicolumn{13}{l}{\it IR carbon stars} \\
MSX\,LMC\,1042     &
                   &
                   &
                   &
                   &
                   &
185                &
                   &
86                 &
PHOT-S             &
639                &
1.44               &
2,5                \\
MSX\,LMC\,1249     &
17.57\rlap{:}      &
14.49              &
10.60              &
                   &
                   &
                   &
                   &
                   &
                   &
                   &
                   &
                   \\
MSX\,LMC\,93       &
                   &
                   &
12.13              &
                   &
                   &
                   &
                   &
                   &
                   &
                   &
                   &
                   \\
MSX\,LMC\,47       &
                   &
                   &
                   &
                   &
                   &
260                &
                   &
67                 &
                   &
707                &
1.79               &
2,5,6              \\
MSX\,LMC\,307      &
                   &
                   &
                   &
                   &
                   &
346                &
                   &
163                &
PHOT-S             &
914                &
1.74               &
2,5,6              \\
MSX\,LMC\,341      &
                   &
                   &
                   &
7                  &
                   &
8                  &
                   &
                   &
                   &
                   &
                   &
1                  \\
MSX\,LMC\,692      &
                   &
14.75              &
10.46              &
                   &
                   &
143                &
                   &
7\rlap{:}          &
                   &
682                &
1.15               &
2,5                \\
MSX\,LMC\,644      &
                   &
                   &
                   &
127                &
                   &
222                &
                   &
                   &
                   &
                   &
                   &
1                  \\
MSX\,LMC\,654      &
                   &
                   &
                   &
118                &
138                &
155                &
135                &
                   &
                   &
                   &
                   &
4                  \\
MSX\,LMC\,872      &
                   &
                   &
                   &
                   &
                   &
171                &
                   &
82                 &
                   &
534                &
1.26               &
2,5,6              \\
\hline
\multicolumn{13}{l}{\it IR objects of unknown type} \\
MSX\,LMC\,733      &
                   &
                   &
                   &
                   &
                   &
525                &
                   &
208                &
CAM-CVF            &
                   &
                   &
2                  \\
\hline
\multicolumn{13}{l}{\it LMC carbon-rich stars from Matsuura et al.\ (2005)} \\
IRAS\,04286$-$6937 &
17.58\rlap{:}      &
15.37              &
10.95              &
                   &
                   &
136                &
                   &
59                 &
                   &
662                &
1.13               &
2,5                \\
IRAS\,04539$-$6821 &
                   &
                   &
                   &
                   &
                   &
244                &
                   &
89                 &
                   &
676                &
1.65               &
2,5                \\
IRAS\,05112$-$6755 &
                   &
                   &
11.65              &
                   &
                   &
387                &
                   &
108                &
PHOT-S             &
826                &
1.72               &
2,5,6              \\
BMB-R\,46          &
14.33              &
11.72              &
9.42               &
67                 &
                   &
51                 &
                   &
                   &
                   &
541                &
0.65               &
1,2,5              \\
\hline
\multicolumn{13}{l}{\it LMC oxygen-rich stars from Matsuura et al.\ (2005)} \\
IRAS\,05128$-$6455 &
17.87              &
13.43              &
                   &
                   &
                   &
226                &
                   &
61                 &
PHOT-S             &
409                &
0.30               &
2,3,5              \\
IRAS\,05148$-$6730 &
10.47              &
                   &
7.34               &
                   &
                   &
380                &
                   &
176                &
                   &
951                &
0.24               &
2,3,5              \\
\hline
\end{tabular}
\end{table*}

\subsection{Photometry}

IR photometry (Table 1) is compiled for all targets and for the Matsuura et
al.\ (2005) objects, except for cluster members (see van Loon et al.\ 2005b).
A few have been monitored at near-IR wavelengths to measure the pulsation
period and amplitude (Table 2).

JKL-band photometry for the targets was obtained at Mount Stromlo and Siding
Springs Observatory (MSSSO) as part of a long-term monitoring campaign (Wood
et al., in preparation). In addition, K$_{\rm s}$-band photometry from the
2-Micron All-Sky Survey (2MASS) is available for almost all targets, but most
targets are too faint in the J-band and many are even too faint in the H-band
to have 2MASS data at those wavelengths. Very few targets are detected in the
DENIS survey, which is inferior to 2MASS in terms of sensitivity. Only three
MSX-selected targets have a DENIS K$_{\rm s}$-band measurement (Table 2).
Additional L$^\prime$-band photometry is available from the spectroscopy
acquisition images, accurate to within a few 0.1 mag.

All prime targets have reliable MSX band A (8.3 $\mu$m) flux densities listed
in the MSX Point Source Catalogue (Version 2.3). We also collected InfraRed
Astronomical Satellite (IRAS) scans from the IRAS data
server\footnote{http://www.astro.rug.nl/IRAS-Server/} and used the Groningen
Image Processing SYstem (GIPSY) software with the {\sc scanaid} tool to
reconstruct a cut through the emission on the exact position. This is superior
to the IRAS Point Source Catalogue values and works well for isolated sources
for which often reliable flux densities can be estimated down to a level of a
few 0.01 Jy at 12 and 25 $\mu$m. A few targets have also been observed with
the Infrared Space Observatory (ISO) (Table 2).

%=========================================================================== 3
\section{Results}

%
% TABLE 3
%
\begin{table*}
\caption[]{Serendipitous objects, listed in order of increasing Right
Ascension (2MASS coordinates, in J2000), together with near-IR magnitudes from
2MASS (JHK$_{\rm s}$) and the spectroscopy acquisition images (L$^\prime$).
The PN candidate and Spitzer Space Telescope target SMP\,LMC\,36 is included
as we obtained L$^\prime$-band photometry for it, although no spectrum was
taken.}
\begin{tabular}{lrrrrrrl}
\hline\hline
Name                                            &
$\alpha$ ($^{\rm h}$ $^{\rm m}$ $^{\rm s}$)     &
$\delta$ ($^\circ$ $^\prime$ $^{\prime\prime}$) &
J                                               &
H                                               &
K$_{\rm s}$                                     &
L$^\prime$                                      &
Type                                            \\
\hline
MSX\,LMC\,50\,B    &
5 02 16.9          &
$-$68 05 24        &
11.25              &
10.40              &
10.17              &
 9.87              &
?                  \\
HV\,11977          &
5 05 45.9          &
$-$68 38 54        &
10.57              &
 9.77              &
 9.53              &
 9.24              &
M giant            \\
{\bf SMP\,LMC\,36}\rlap{$^a$} &
5 10 39.7          &
$-$68 36 05        &
16.25              &
15.61              &
14.00              &
11.24              &
Planetary Nebula?  \\
MSX\,LMC\,349\,B   &
5 17 24.4          &
$-$68 55 06        &
11.62              &
10.50              &
 9.83              &
 9.98              &
Carbon star        \\
HV\,2532           &
5 26 27.4          &
$-$69 10 56        &
 8.93              &
 8.08              &
 7.71              &
 7.42              &
M4 supergiant      \\
MSX\,LMC\,1780\,B  &
5 32 59.0          &
$-$68 23 51        &
11.10              &
10.54              &
10.46              &
10.12              &
Early-type star?   \\
\hline
\end{tabular}\\
Spitzer programme: $a=$ GTO \#103 (P.I.: J.\ Houck).
\end{table*}

%------------------------------------------------------------------------- 3.1
\subsection{IR carbon stars}

The majority (28) of our targets turn out to be carbon stars, on the basis of
strong carbonaceous molecular absorption features seen in the spectrum around
3.1 and often 3.8 $\mu$m (Fig.\ 2). Out of these, 10 are part of Spitzer Space
Telescope GO programme \#3505 (PI: P.\ Wood).

It is worth noting that MSX\,LMC\,307 has the longest known pulsation period
of any carbon star: 889 d (Wood 1998) or 939 d (Whitelock et al.\ 2003).

MSX\,LMC\,55 is misidentified in the literature with the M-type giant
HV\,11977, of which we also took a spectrum.

Another mistaken identity occurs for MSX\,LMC\,45: Leisy et al.\ (1997)
identified the IRAS source with the low-excitation Planetary Nebula
SMP\,LMC\,36 (Sanduleak, MacConnell \& Philip 1978), located at
$9^{\prime\prime}$ WNW from an extremely red 2MASS source. Our 3--4 $\mu$m
spectrum is of the 2MASS object, whilst the PN is a target for Spitzer Space
Telescope GTO programme \#103 (P.I.: J.\ Houck). Although the K$_{\rm
s}$--L$^\prime$ colour of SMP\,LMC\,36 (Table 3) is very red at 2.76 mag,
MSX\,LMC\,45 is much redder at 4.95 mag and with $L^\prime=7.87$ mag much
brighter in the thermal IR than SMP\,LMC\,36 which has $L^\prime=11.24$ mag.
For all effects the emission detected at 8.3 (MSX), 12 and 25 $\mu$m (IRAS)
arises from the carbon star.

%------------------------------------------------------------------------- 3.2
\subsection{IR objects of unknown type}

For two of our targets, the 3--4\,$\mu$m spectrum does not lead to an
unambiguous photospheric chemical classification (Fig.\ 3). Coincidentally,
these are the only two objects in our target sample that are part of Spitzer
Space Telescope programmes other than GO \#3505. Three more objects turned out
to be massive Young Stellar Objects; these are analysed in detail in van Loon
et al.\ (2005c) and Oliveira et al.\ (in preparation).

The 3--4 $\mu$m spectrum of MSX\,LMC\,196 displays a hint of absorption around
3.0--3.1 and 3.7--3.8 $\mu$m, in which case it is a carbon star. The molecular
features may be heavily veiled by the strong dust continuum emission. Indeed,
with $F_{12}=F_{25}$ its [12]--[25] colour is redder than any of the IR carbon
stars.

On the basis of the IR colours, van Loon et al.\ (1998) classified the dust
around MSX\,LMC\,733 as oxygen-rich, but in a re-analysis including ISO data
Trams et al.\ (1999) classified it as carbon-rich. They also present a CAM-CVF
spectrum, in which broad emission around 11.3 $\mu$m was interpreted as SiC,
which would confirm its carbon-rich nature. However, its CAM-CVF spectrum is
not unlike that of OH/IR stars where the silicate feature is seen in
self-absorption. Overall the 3--4 $\mu$m spectrum of MSX\,LMC\,733 is similar
to the ISO-SWS spectrum of the M8 giant CD\,Gru (Vandenbussche et al.\ 2002).

%------------------------------------------------------------------------- 3.3
\subsection{Serendipitous objects}

In several cases a spectrum could be taken of another IR-bright object,
simultaneously with the principal target. These ``serendipitous'' objects
(Fig.\ 4 \& Table 3) are not generally dust-enshrouded.

The M4 supergiant HV\,2532 is responsible for most of the mid-IR emission
associated with IRAS\,05267$-$6913 ($F_{12}\sim0.20$ Jy, $F_{25}\sim0.25$ Jy).
The MSX data clearly separates its emission, MSX\,LMC\,592 ($F_{\rm A}=230$
mJy), from that of our spectroscopy target, MSX\,LMC\,774.

%------------------------------------------------------------------------- 3.4
\subsection{Cluster IR objects}

We took spectra of six cluster IR objects from van Loon et al.\ (2005b) (Fig.\
5). Four of these, NGC\,1903-IR1 and IR2 and NGC\,1978-IR1 and IR2 are carbon
stars. The other two, NGC\,1984-IR1 and IR2 are oxygen-rich objects; IR1
(=IRAS\,05280$-$6910) is a source of OH (Wood et al.\ 1992; Marshall et al.\
2004) and H$_2$O maser emission (van Loon et al.\ 2001) whilst IR2 is the M1
supergiant WOH G347.

Two objects in our MSX-selected target list, MSX\,LMC\,484 and 654, and three
of the carbon stars with 3--4 $\mu$m spectra published in Matsuura et al.\
(2005), IRAS\,04496$-$6958, IRAS\,04557$-$6753 and LI-LMC\,1813 were also
found to be associated with clusters (van Loon et al.\ 2003; van Loon et al.\
2005b).

%
% FIGURE 2
%
\begin{figure*}[tb]
\centerline{\psfig{figure=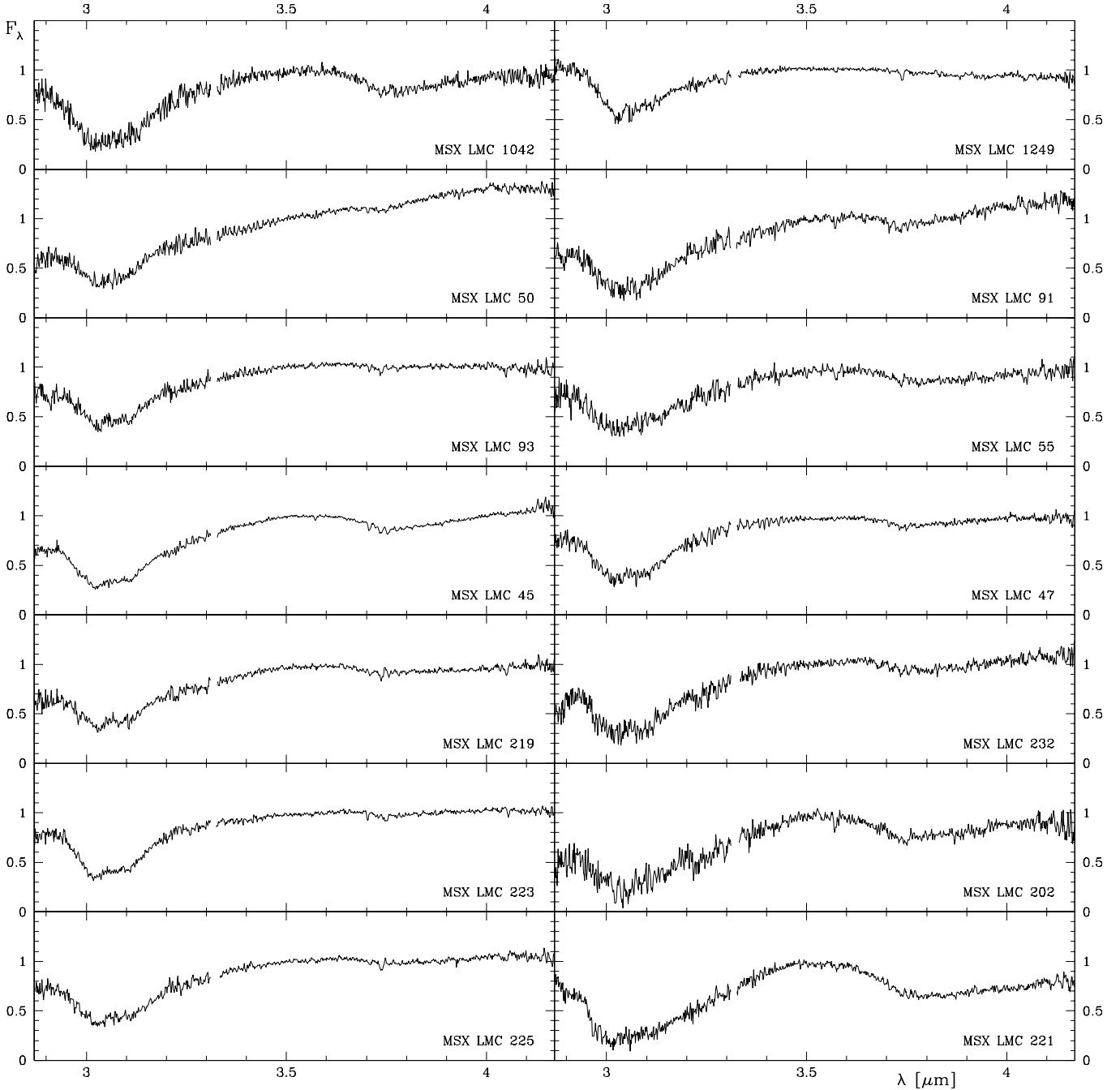,width=180mm}}
\caption[]{ESO/VLT 3--4 $\mu$m spectra of MSX-selected IR carbon stars. The
spectra have been normalised to $F_{3.5}=1$.}
\end{figure*}

%
% FIGURE 2 (continued)
%
\begin{figure*}[tb]
\addtocounter{figure}{-1}
\centerline{\psfig{figure=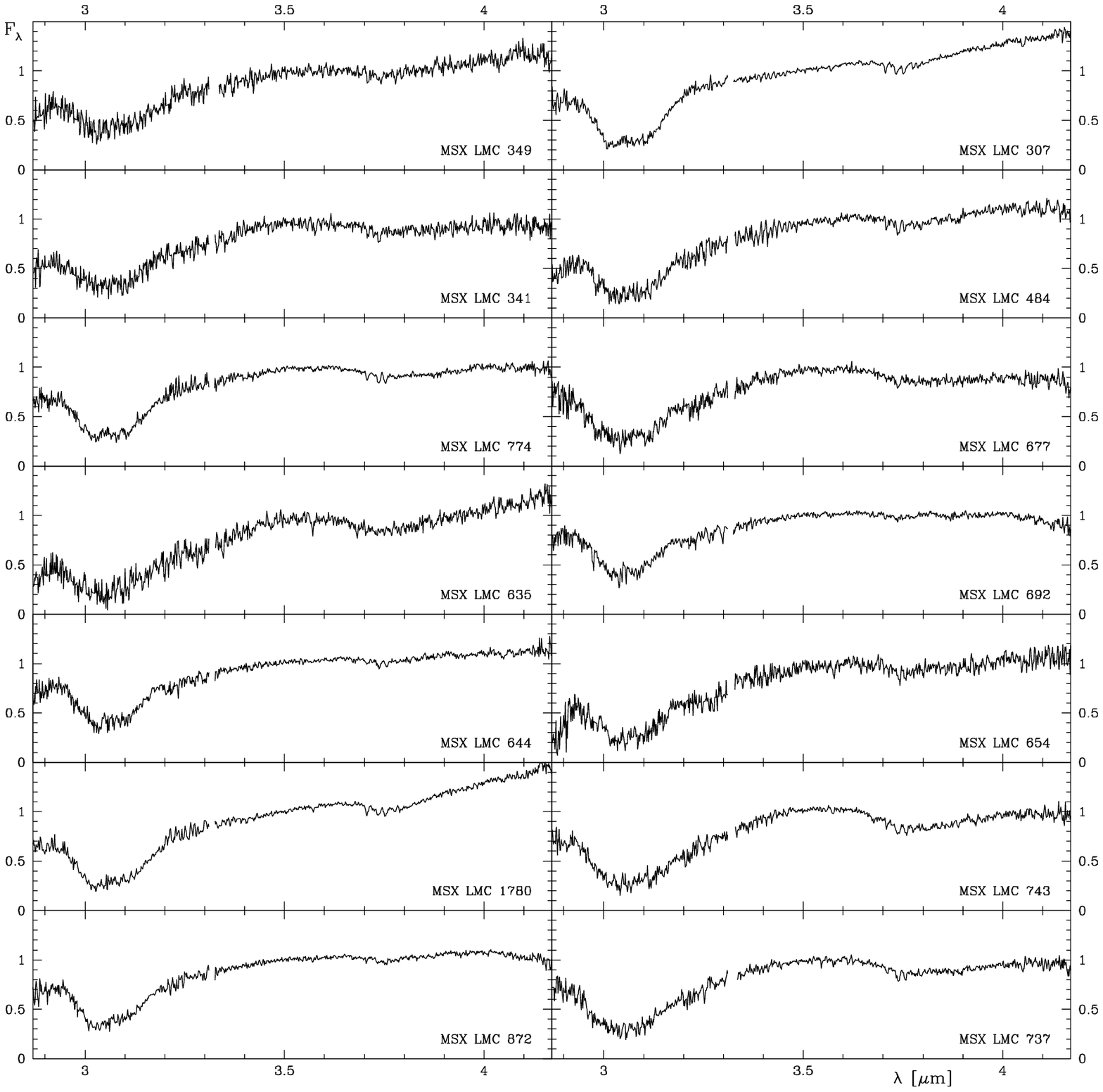,width=180mm}}
\caption[]{(Continued)}
\end{figure*}

%
% FIGURE 3
%
\begin{figure}[tb]
\centerline{\psfig{figure=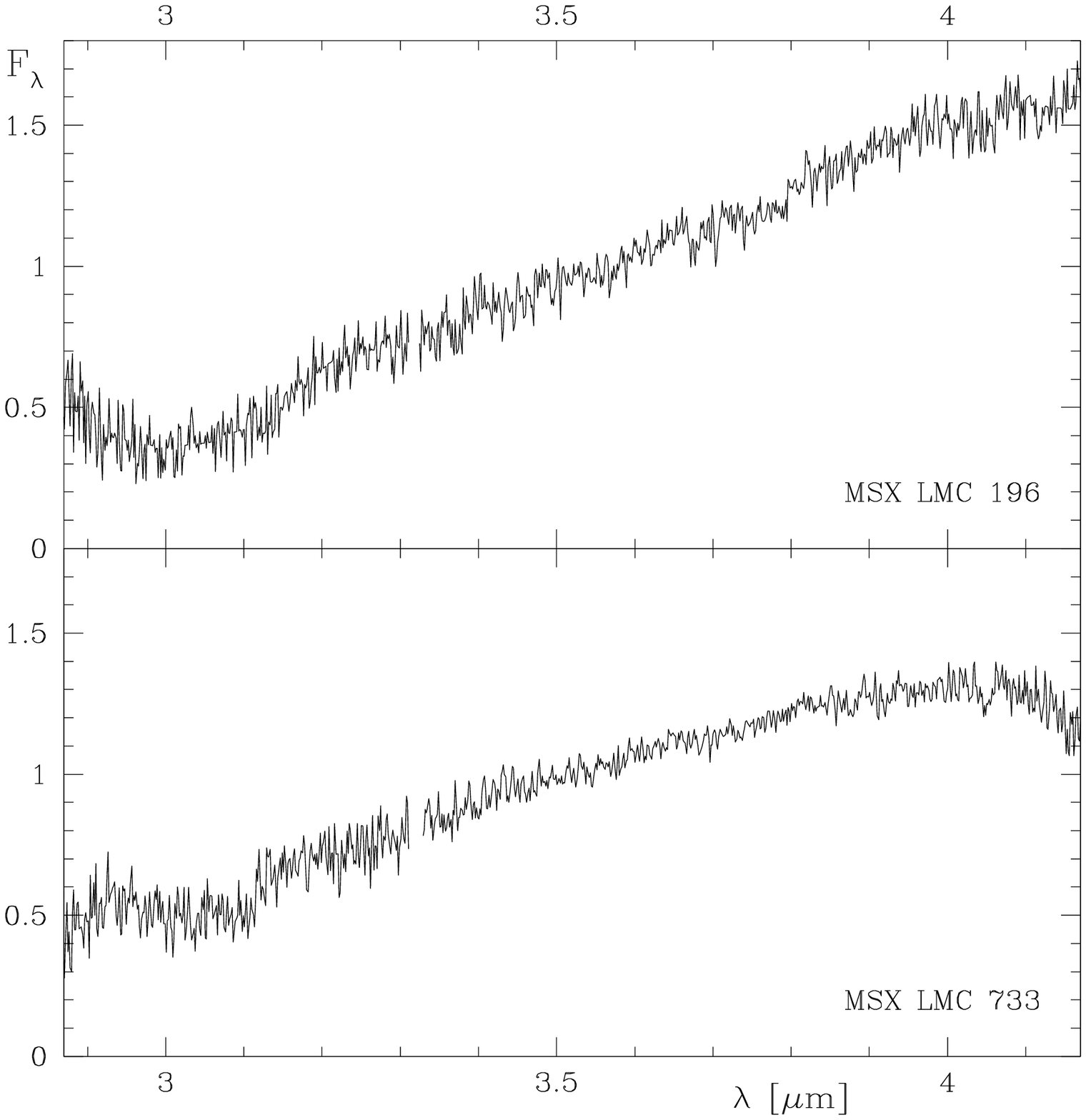,width=88mm}}
\caption[]{ESO/VLT 3--4 $\mu$m spectra of IR objects of unknown type.
MSX\,LMC\,196 shows a hint of absorption around 3.1 $\mu$m and 3.65--3.80
$\mu$m, and may thus be a heavily veiled carbon star. The overall shape of the
spectrum of MSX\,LMC\,733 resembles that of cool oxygen-rich stars better.}
\end{figure}

%
% FIGURE 4
%
\begin{figure}[tb]
\centerline{\psfig{figure=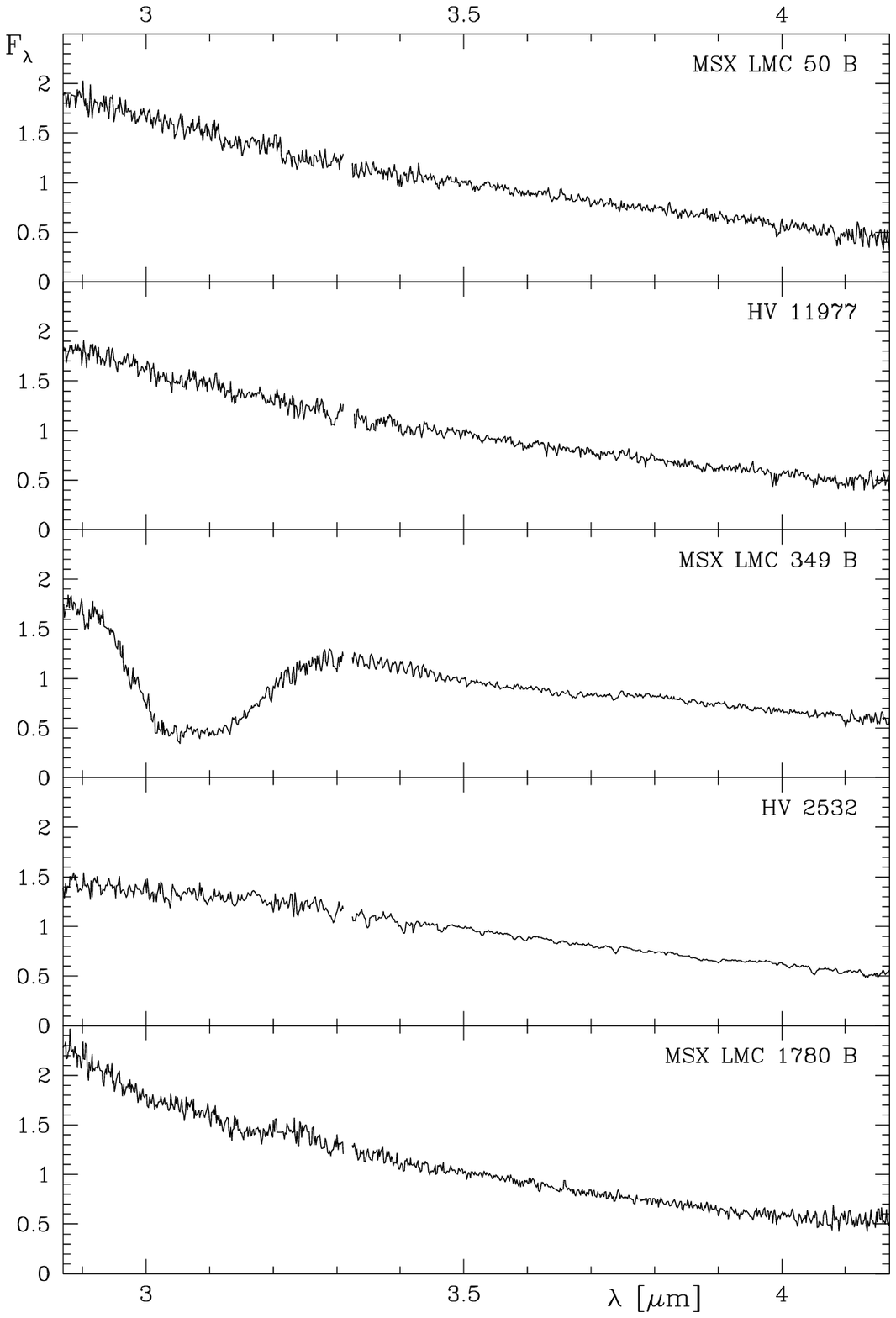,width=88mm}}
\caption[]{Serendipitous ESO/VLT 3--4 $\mu$m spectra (see Table 3). They are
all oxygen-rich except MSX\,LMC\,349\,B which is an optically bright (yet
anonymous) carbon star.}
\end{figure}

%=========================================================================== 4
\section{Analysis}

%------------------------------------------------------------------------- 4.1
\subsection{Luminosities and mass-loss rates}

As a result of the selection criteria, the MSX-selected targets have very red
K$_{\rm s}$--L$^\prime$ (Fig.\ 6). This is caused by circumstellar extinction
diminishing the K$_{\rm s}$-band brightness and circumstellar dust emission
enhancing the L$^\prime$-band brightness. Hence their faint K$_{\rm s}$-band
brightness is not a direct consequence of their bolometric luminosity. IR
carbon stars are objects which are known to be variable with amplitudes up to
2 magnitudes at near-IR wavelengths (e.g., Wood 1998; Whitelock et al.\ 2003).
Although the 2MASS, MSX and IRAS data refer to combined measurements taken at
different epochs, the groundbased near-IR data that we use here are
single-epoch measurements.

The spectral energy distributions of the IR carbon stars were modelled with
the radiative transfer code {\sc dusty} (Ivezi\'{c}, Nenkova \& Elitzur 1999).
The density distribution is based upon a hydrodynamic computation of a
dust-driven wind at constant mass-loss rate. The model was then scaled to
match the overall observed Spectral Energy Distribution (SED), knowing the
distance to the LMC (50 kpc), which then yields an accurate measurement of the
bolometric luminosity.

A blackbody was used to represent the underlying stellar radiation field, with
a temperature of $T_{\rm eff}=2500$ K. We used 0.1 $\mu$m grains of amorphous
carbon dust (Henning \& Mutschke 1997). To obtain the mass-loss rate one has
to know the dust grain density, for which we adopt $\rho_{\rm grain}=3$ g
cm$^{-3}$, and the gas-to-dust mass ratio, for which we adopt $\rho_{\rm
gas}/\rho_{\rm dust}=500$ at the metallicity of the LMC (two to three times
below solar). The results are summarised in Table 4.

We usually found that a satisfactory match between the model and the observed
SED required a rather low temperature for the dust at the inner edge of the
envelope, $T_{\rm dust}\sim600$ to 700 K. The derived bolometric luminosity
and mass-loss rate do not however depend sensitively on the inner-edge dust
temperature. The estimated internal accuracies of the luminosity and mass-loss
rate are about 5 and 10 per cent, respectively.

The mass-loss rates are correlated with the bolometric luminosities (Fig.\ 7).
Most stars straddle the classical limit to the mass-loss rate for single
scattering (long-dashed line in Fig.\ 7; van Loon et al.\ 1999b). The observed
sequence is steeper than the classical limit though, with luminous stars
experiencing mass-loss rates that are higher than this limit --- but not as
high as the maximum mass-loss rates observed in the LMC (short-dashed lines in
Fig.\ 7; van Loon et al.\ 1999b). The two carbon stars with particularly long
pulsation periods, $P>800$ d, and amplitudes, ${\Delta}K_{\rm s}>1.7$ mag,
MSX\,LMC\,307 and IRAS\,05112$-$6755 (Table 2) are amongst the stars with the
highest mass-loss rates and luminosities.

%
% FIGURE 5
%
\begin{figure*}[tb]
\centerline{\psfig{figure=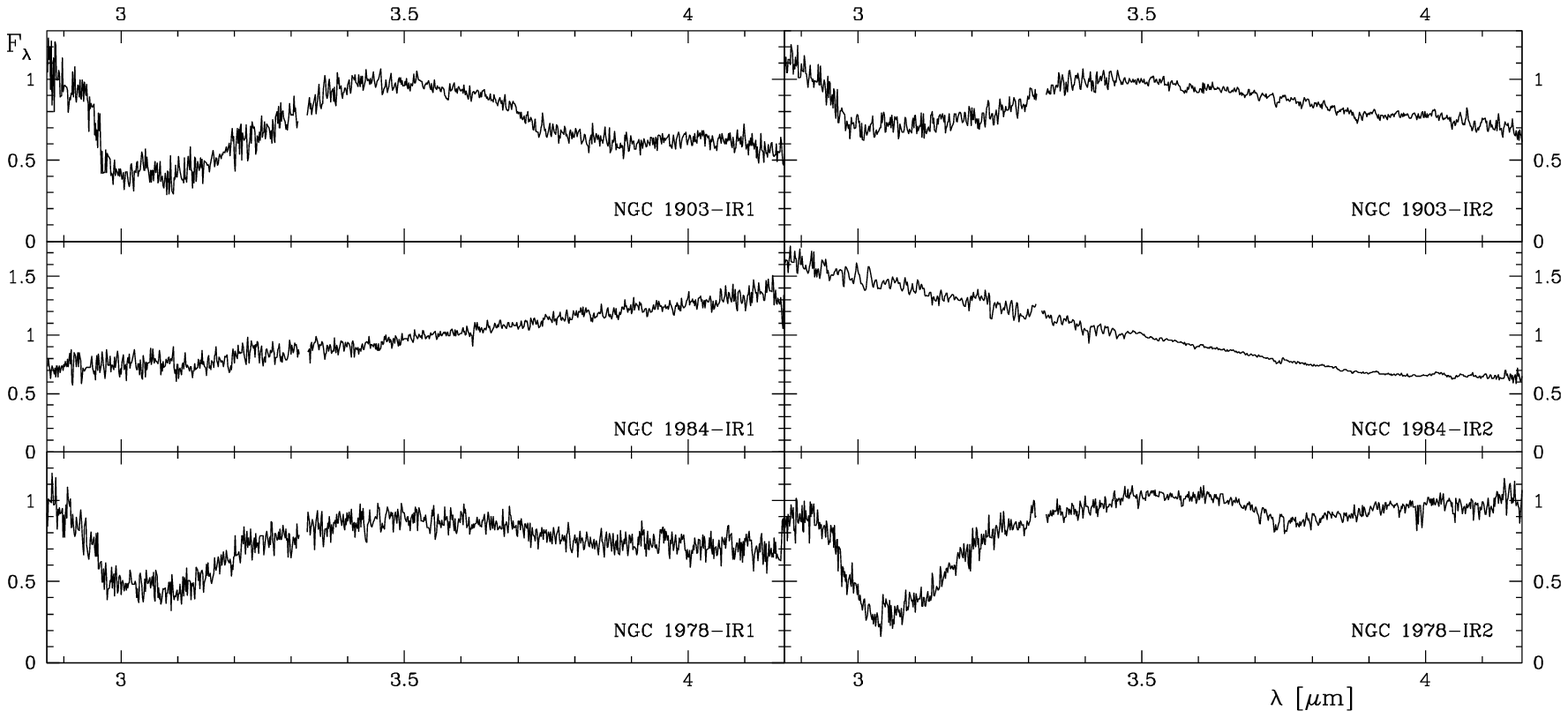,width=180mm}}
\caption[]{ESO/VLT 3--4 $\mu$m spectra of additional objects in or near
populous star clusters (van Loon et al.\ 2005b). They are all carbon stars
except NGC\,1984-IR1 and -IR2 which are an OH/IR star and an optically bright
red supergiant, respectively.}
\end{figure*}

%
% FIGURE 6
%
\begin{figure}[tb]
\centerline{\psfig{figure=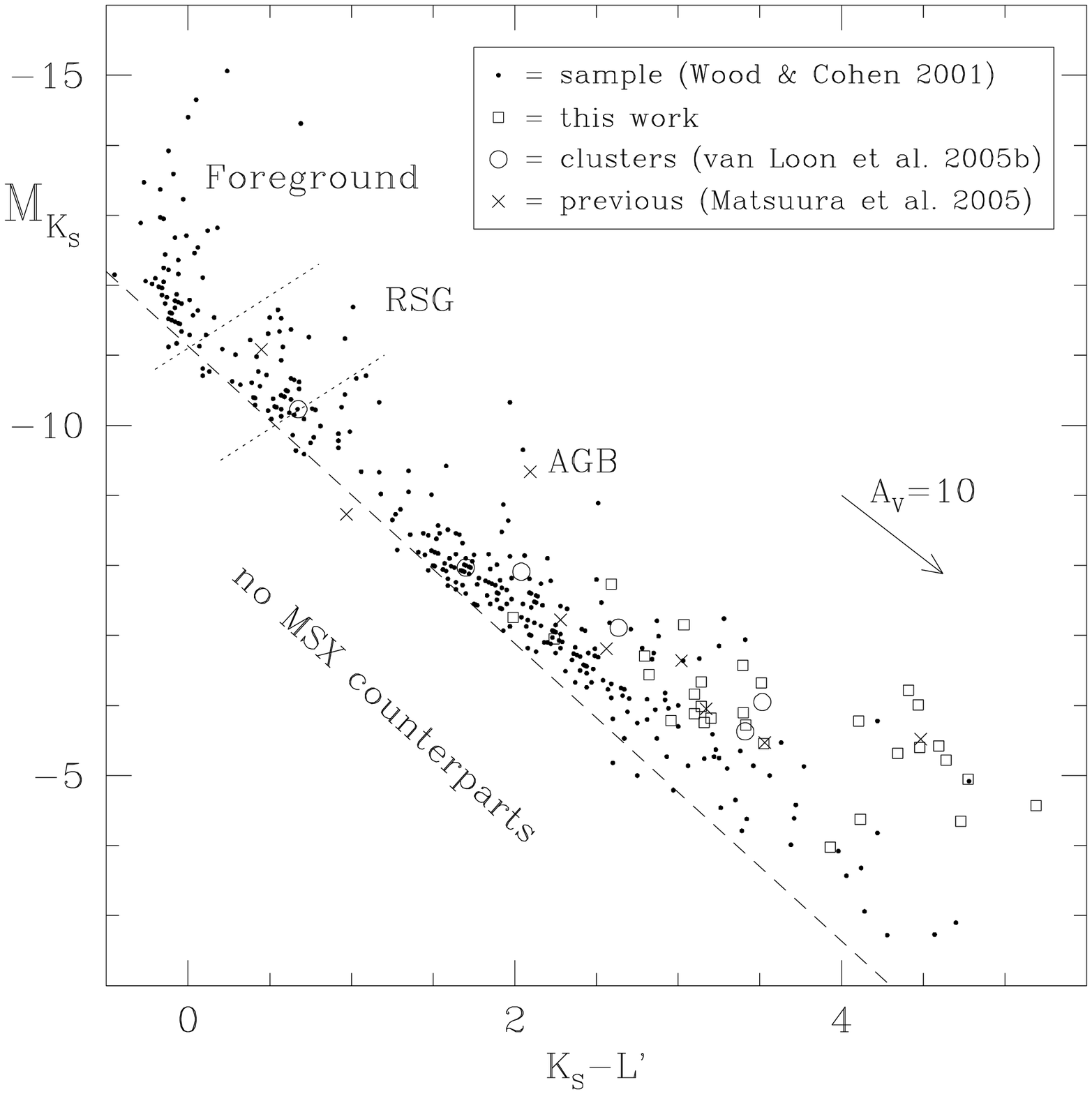,width=88mm}}
\caption[]{Absolute K$_{\rm s}$ brightness versus K$_{\rm s}$--L$^\prime$
colour-magnitude diagram of the MSX sample, where the objects for which we
present spectra (Table 1) are indicated by squares and circles.}
\end{figure}

%
% TABLE 4
%
\begin{table}
\caption[]{Bolometric luminosities, mass-loss rates, inner-edge dust
temperatures and optical depth of the IR carbon stars. They are grouped (last
column) as follows: 1=high $\dot{\rm M}$ and L; 2=moderate $\dot{\rm M}$ and
L; 3=low $\dot{\rm M}$ and L; 4=low $\dot{\rm M}$ but high L.}
\begin{tabular}{lcccrc}
\hline\hline
Name                                &
$\log({\rm L})$                     &
$\log(\dot{\rm M})$                 &
T$_{\rm dust}$                      &
$\tau_{\rm V}$                      &
G                                   \\
                                    &
(L$_\odot$)                         &
\llap{(M}$_\odot$ yr\rlap{$^{-1}$)} &
(K)                                 &
                                    &
                                    \\
\hline
\multicolumn{6}{l}{\it IR carbon stars} \\
MSX\,LMC\,1042     &
3.91               &
$-4.80$            &
600                &
13                 &
3                  \\
MSX\,LMC\,1249     &
4.07               &
$-4.84$            &
700                &
10                 &
4                  \\
MSX\,LMC\,50       &
3.95               &
$-4.51$            &
650                &
33                 &
2                  \\
MSX\,LMC\,91       &
3.85               &
$-4.77$            &
650                &
19                 &
3                  \\
MSX\,LMC\,93       &
3.88               &
$-4.82$            &
700                &
17                 &
3                  \\
MSX\,LMC\,55       &
4.04               &
$-4.74$            &
450                &
4                  &
4                  \\
MSX\,LMC\,45       &
4.31               &
$-4.30$            &
650                &
28                 &
1                  \\
MSX\,LMC\,47       &
4.07               &
$-4.51$            &
650                &
25                 &
2                  \\
MSX\,LMC\,219      &
4.03               &
$-4.58$            &
700                &
25                 &
2                  \\
MSX\,LMC\,232      &
3.90               &
$-4.73$            &
600                &
17                 &
3                  \\
MSX\,LMC\,223      &
4.00               &
$-4.54$            &
700                &
30                 &
2                  \\
MSX\,LMC\,202      &
3.78               &
$-4.76$            &
700                &
25                 &
3                  \\
MSX\,LMC\,225      &
4.01               &
$-4.48$            &
650                &
32                 &
2                  \\
MSX\,LMC\,221      &
3.94               &
$-4.65$            &
550                &
17                 &
2                  \\
MSX\,LMC\,349      &
3.90               &
$-4.54$            &
700                &
37                 &
2                  \\
MSX\,LMC\,307      &
4.15               &
$-4.38$            &
650                &
31                 &
1                  \\
MSX\,LMC\,341      &
3.80               &
$-4.75$            &
700                &
25                 &
3                  \\
MSX\,LMC\,484      &
3.71               &
$-4.78$            &
\llap{1}000        &
31                 &
3                  \\
MSX\,LMC\,774      &
4.00               &
$-4.60$            &
600                &
20                 &
2                  \\
MSX\,LMC\,677      &
3.88               &
$-4.85$            &
700                &
15                 &
3                  \\
MSX\,LMC\,635      &
4.27               &
$-4.31$            &
650                &
29                 &
1                  \\
MSX\,LMC\,692      &
4.01               &
$-4.70$            &
700                &
18                 &
4                  \\
MSX\,LMC\,644      &
4.02               &
$-4.71$            &
700                &
17                 &
4                  \\
MSX\,LMC\,654      &
3.88               &
$-4.67$            &
700                &
20                 &
2                  \\
MSX\,LMC\,1780     &
4.07               &
$-4.35$            &
600                &
37                 &
1                  \\
MSX\,LMC\,743      &
3.76               &
$-4.85$            &
700                &
20                 &
3                  \\
MSX\,LMC\,872      &
3.94               &
$-4.61$            &
700                &
28                 &
2                  \\
MSX\,LMC\,737      &
4.07               &
$-4.69$            &
700                &
16                 &
4                  \\
\hline
\multicolumn{6}{l}{\it Cluster carbon stars from van Loon et al.\ (2005b)} \\
NGC\,1903-IR1      &
3.99               &
$-$5.11            &
\llap{1}300        &
9                  &
                   \\
NGC\,1903-IR2      &
3.84               &
$-$5.43            &
\llap{1}500        &
6                  &
                   \\
NGC\,1978-IR1      &
3.73               &
$-$5.05            &
\llap{1}000        &
9                  &
                   \\
NGC\,1978-IR2      &
3.77               &
$-$4.80            &
\llap{1}100        &
22                 &
                   \\
\hline
\multicolumn{6}{l}{\it LMC carbon-rich stars from Matsuura et al.\ (2005)} \\
IRAS\,04286$-$6937 &
3.95               &
$-$4.43            &
420                &
15                 &
                   \\
IRAS\,04496$-$6958 &
4.61               &
$-$4.54            &
720                &
6                  &
                   \\
IRAS\,04539$-$6821 &
3.94               &
$-$4.54            &
700                &
27                 &
                   \\
IRAS\,04557$-$6753 &
3.99               &
$-$4.57            &
600                &
16                 &
                   \\
IRAS\,05112$-$6755 &
4.22               &
$-$4.25            &
500                &
20                 &
                   \\
BMB-R\,46          &
4.19               &
$-$5.27            &
700                &
1                  &
                   \\
LI-LMC\,1813       &
4.09               &
$-$4.33            &
600                &
28                 &
                   \\
\hline
\end{tabular}
\end{table}

On the whole the mass-loss rates of the IR carbon stars are higher than those
of optically bright carbon stars (van Loon et al.\ 2005a), but amongst the IR
carbon stars themselves are stars that have relatively low mass-loss rates for
their high luminosity. Four of the carbon stars from van Loon et al.\ (2005b)
and Matsuura et al.\ (2005) do not appear in Fig.\ 7 as they have
significantly lower mass-loss rates than the stars in the MSX-selected sample.
These stars may be in a slightly different evolutionary phase, for instance if
they have not yet entered the superwind stage or if they pass through a more
quiet episode in their thermal-pulsing cycle. Differences in mass may also
contribute to a spread in the mass-loss rate-luminosity plane --- see van Loon
et al.\ (2005b) for a discussion of the mass-loss rates and luminosities of
cluster IR objects in the Magellanic Clouds. The by far most luminous carbon
star that we analyse here, IRAS\,04496$-$6958 also lies beyond the boundaries
of the diagram. Its mass-loss rate is fairly modest for its high luminosity.
In this case there are strong indications for a relatively massive progenitor
star ($M_{\rm initial}\sim5$ M$_\odot$) that may have experienced a thermal
pulse recently (van Loon et al.\ 2005b). For the purpose of the spectral
analysis, we divide the sample of IR carbon stars into four groups on the
basis of their mass-loss rate and luminosity (Fig.\ 7 \& Table 4).

Stars with high mass-loss rates have a red K$_{\rm s}$--L$^\prime$ colour or,
if having a relatively modestly red K$_{\rm s}$--L$^\prime$ colour they have a
red L$^\prime$--[8.3] colour (Fig.\ 8). The fact that the L$^\prime$--[8.3]
colour shows some correlation with the K$_{\rm s}$--L$^\prime$ colour but with
a large scatter implies that besides the optical depth of the dust shell other
parameters play a r\^{o}le. These may include the dust temperature at the
inner edge of the envelope and perhaps the dust type, or photometric
variability.

%
% FIGURE 7
%
\begin{figure}[tb]
\centerline{\psfig{figure=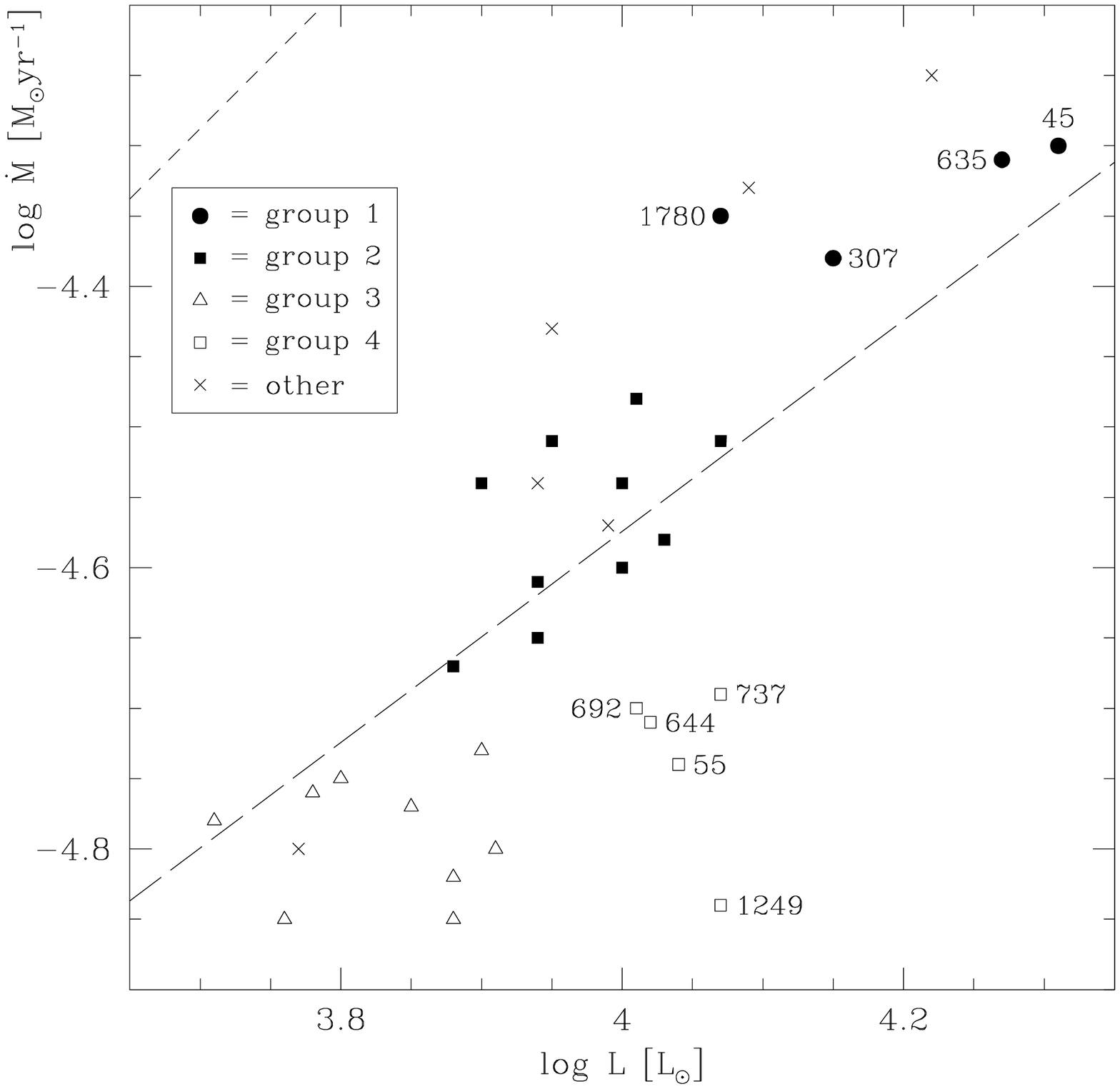,width=88mm}}
\caption[]{Mass-loss rate versus luminosity diagram for IR carbon stars, which
we use to subdivide the sample into four groups represented by different
symbols. Stars from the cluster sample of van Loon et al.\ (2005b) and
previously published spectra of Matsuura et al.\ (2005) are kept separate
(crosses).}
\end{figure}

%
% FIGURE 8
%
\begin{figure}[tb]
\centerline{\psfig{figure=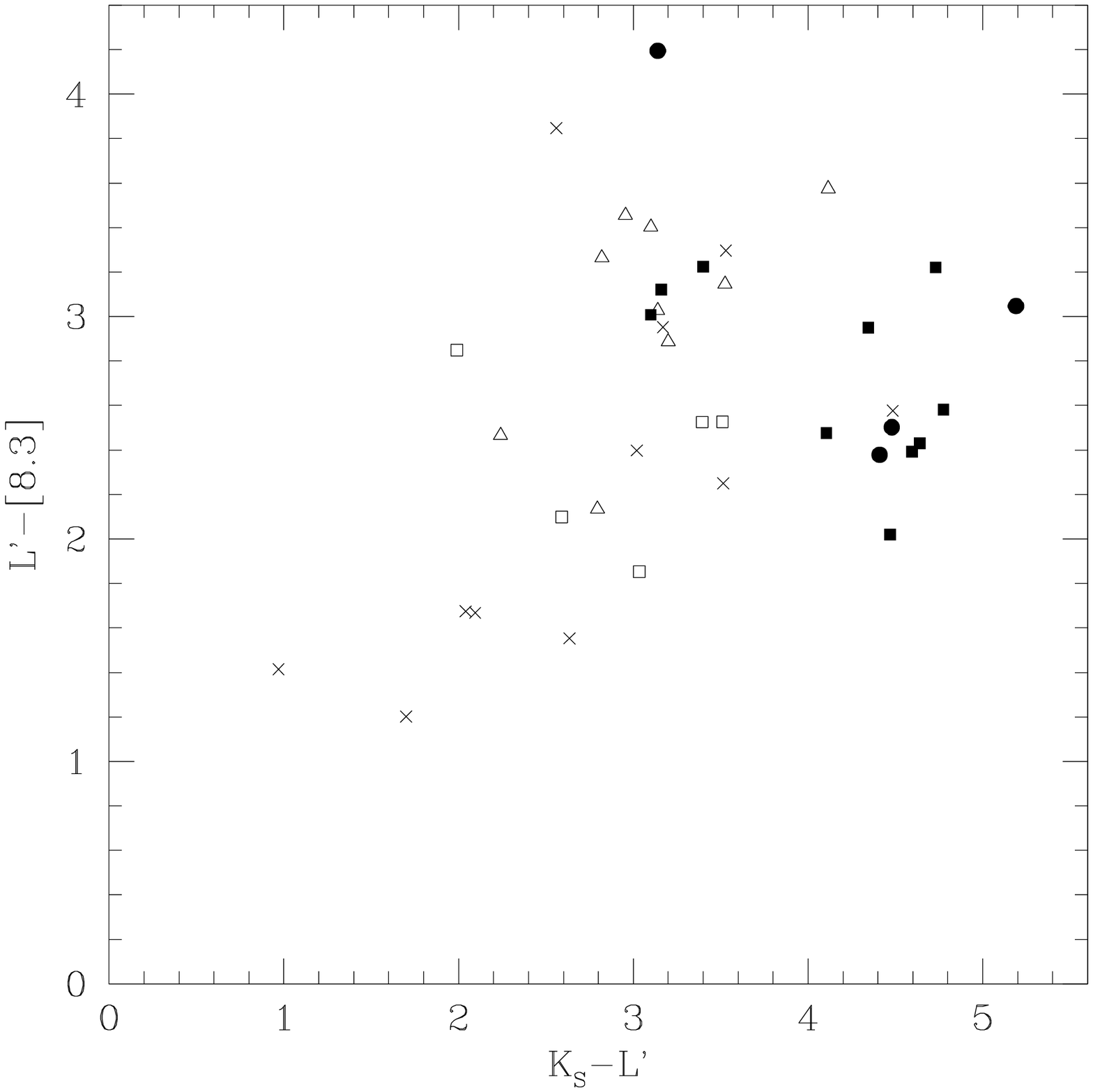,width=88mm}}
\caption[]{L$^\prime$--[8.3] versus K$_{\rm s}$--L$^\prime$ colour-colour
diagram of the IR carbon stars. Redder colours are generally indicative of a
larger optical depth of the circumstellar dust envelope, but there is
significant scatter due to other (circum)stellar parameters and
non-contemporaneous photometry. Symbols are as in Fig.\ 7.}
\end{figure}

%------------------------------------------------------------------------- 4.2
\subsection{Identification of spectral features}

A grand average of all spectra of the IR carbon stars (Fig.\ 9) shows all the
spectral features of interest at a much reduced noise level. Examples of four
types of spectral appearance are given in Fig.\ 10, to show the similarities
and differences amongst the IR carbon stars.

Hydrogen absorption lines of the Brackett and Pfund series are seen in many of
the spectra, but this may be due to residuals in the correction for the same
lines in the telluric standard star spectrum. There is no evidence in any of
the carbon stars for strong hydrogen line emission arising from shocks in
their pulsating atmospheres. A few spectra show an array artifact causing a
sharp feature in the extracted spectrum around 3.99 $\mu$m.

The strongest absorption band, between about 2.95 and 3.15 $\mu$m, is seen in
all IR carbon stars, and is due to a blend of HCN and C$_2$H$_2$. The shape is
not universal, although often with characteristic peaks at 3.01 and 3.11
$\mu$m it sometimes has a rather triangular shape peaking at 3.01 $\mu$m only
(e.g., MSX\,LMC\,1249). Often, the absorption seems to extend to as far as 3.5
$\mu$m (e.g., MSX\,LMC\,221), but it is not certain that this is still part of
the same band. Alternatively, the HNC isomer of HCN is expected to contribute
in precisely this spectral region (Harris et al.\ 2003).

Most IR carbon stars exhibit a broad depression in the spectrum around 3.8
$\mu$m. In the LMC, the absorption almost always peaks around 3.75 $\mu$m,
whereas in galactic carbon stars absorption is normally seen around 3.9 $\mu$m
instead. In our sample, only the less extreme object NGC\,1903-IR1 shows
evidence for distinct absorption at 3.9 $\mu$m. CH was suggested to explain a
series of absorption lines around 3.75 $\mu$m in the spectra of the galactic
carbon stars TX\,Psc and WZ\,Cas (Aoki, Tsuji \& Ohnaka 1998), but the CH does
not lead to a broad absorption band. Instead, the identification with
C$_2$H$_2$ by Cernicharo et al.\ (1999) matches the broad absorption seen
around 3.75 $\mu$m in the galactic dust-enshrouded carbon star IRC\,+10216.
HCN is held responsible for the 3.9 $\mu$m component (Aoki et al.\ 1998;
Harris et al.\ 2003), but CS may also contribute (Aoki et al.\ 1998).

A few quite narrow absorption features are seen, superimposed onto the 3.8
$\mu$m absorption band (Fig.\ 9). The 3.70 $\mu$m feature is the most
conspicuous of these, but it is always seen in conjunction with a similar
feature at 3.76 $\mu$m. A weaker feature at 3.78 $\mu$m may be related, and
there seems to be another feature that is however blended with the Pf$\gamma$
line at 3.74 $\mu$m. Aoki et al.\ (1998) interpret 3.70 and 3.77 $\mu$m
features as being part of the CH(3-2) fundamental band, but the stronger
associated 3.64 and 3.67 $\mu$m features (cf.\ J{\o}rgensen et al.\ 1996) are
not seen in our spectra (whilst they are seen in TX\,Psc and WZ\,Cas) ---
except perhaps in the enigmatic object MSX\,LMC\,196 (Fig.\ 3). The features
might be part of the C$_2$H$_2$ band instead (Cernicharo et al.\ 1999).

%
% FIGURE 9
%
\begin{figure*}[tb]
\centerline{\psfig{figure=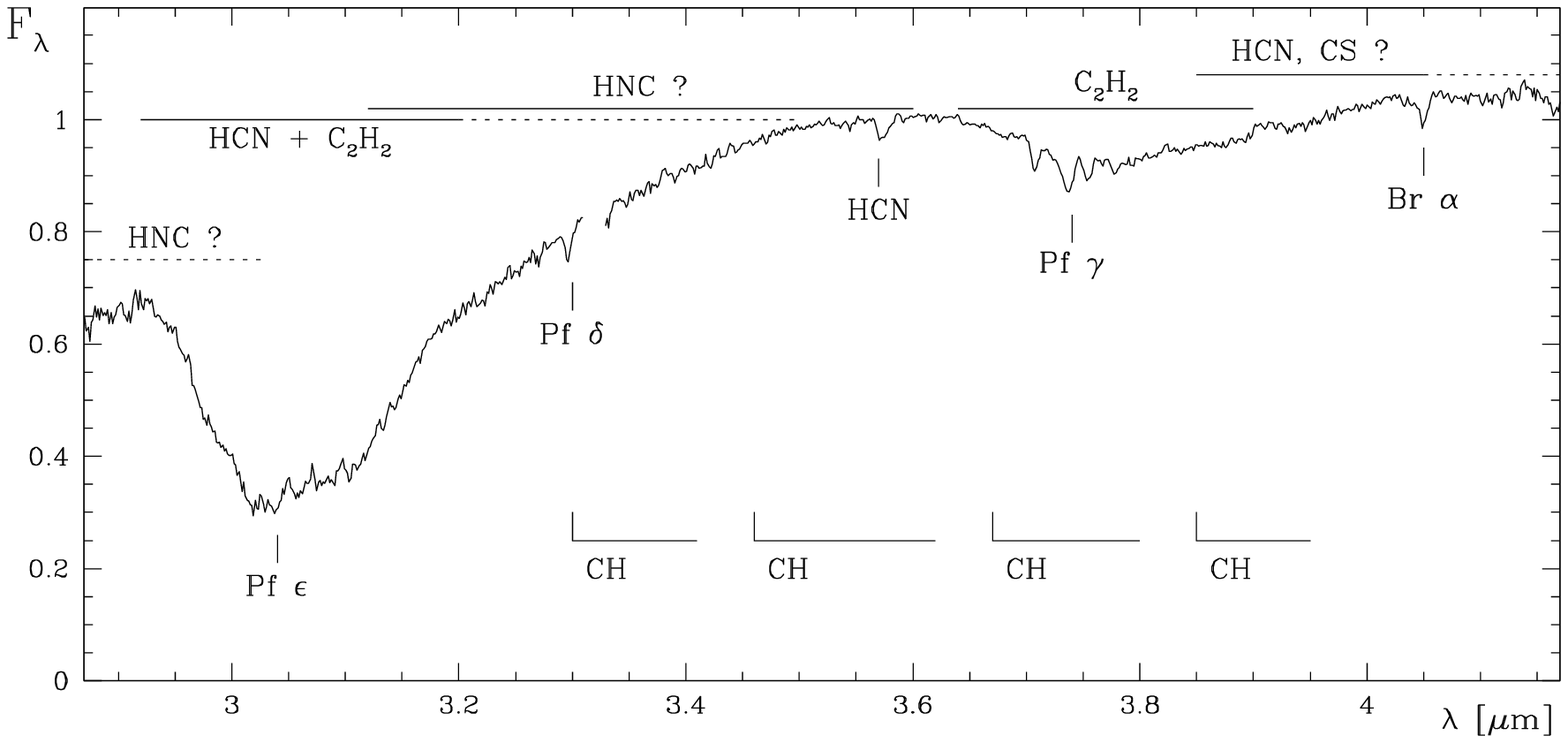,width=170mm}}
\caption[]{Grand average of all IR carbon star spectra (Table 1), with
identifications of the main features discussed in the text.}
\end{figure*}

CH(1-0) may be responsible for the series of narrow absorption lines seen
around 3.3--3.5 $\mu$m in the spectrum of MSX\,LMC\,349\,B (Fig.\ 4). The
spectrum of this optically bright carbon star is in fact slightly depressed up
to a wavelength of about 3.75 $\mu$m.

A comparison between the average spectra for the four groups in terms of
mass-loss rate and luminosity (Fig.\ 11) fails to reveal any striking
correlation of the spectral features with either of these parameters. The
dispersion within each group average is quite large, but the error in the mean
difference between the group average and the grand average indicates
statistically significant differences between the groups. The most obvious
difference is in the slope of the spectral continuum, which depends on the
optical depth of the circumstellar dust envelope: the continuum is redder at
higher mass-loss rates, but at the same mass-loss rate it is redder at lower
luminosity.

%
% FIGURE 10
%
\begin{figure}[tb]
\centerline{\psfig{figure=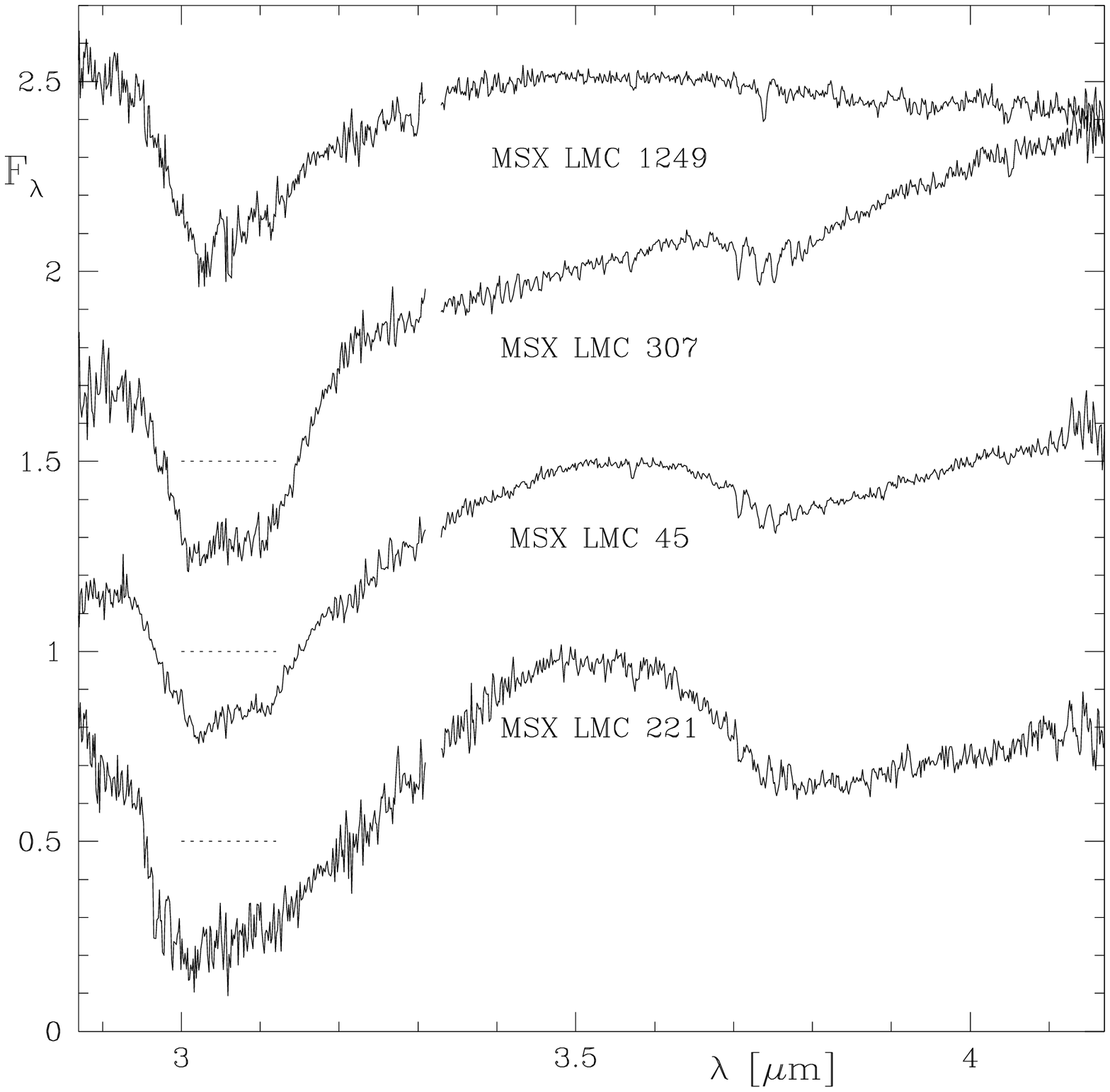,width=88mm}}
\caption[]{Examples of four types of spectral appearance amongst the IR carbon
stars.}
\end{figure}

%
% FIGURE 11
%
\begin{figure}[tb]
\centerline{\psfig{figure=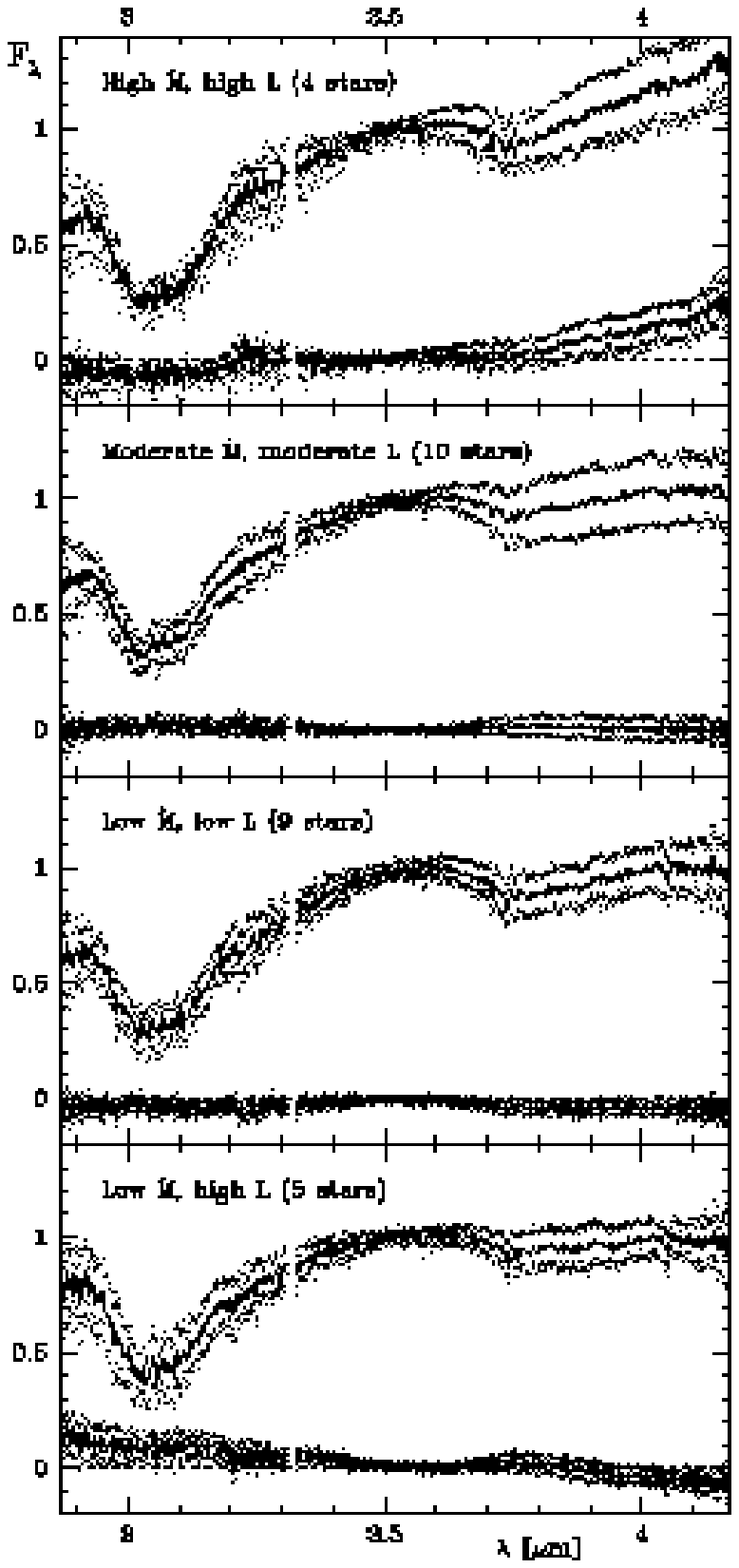,width=88mm}}
\caption[]{Average spectra for the four groups of stars selected on the basis
of their mass-loss rate and luminosity, plotted together with their difference
spectra with respect to the grand average of all IR carbon star spectra (see
Fig.\ 9). The dots indicate the dispersion in the group averages and the error
in the mean difference between the group and grand averages.}
\end{figure}

%------------------------------------------------------------------------- 4.3
\subsection{Molecular band strengths}

%
% TABLE 5
%
\begin{table}
\caption[]{The band and blue/red continuum definitions for the computation of
equivalent widths. Wavelengths are in $\mu$m.}
\begin{tabular}{lccc}
\hline\hline
Feature        &
band           &
blue continuum &
red continuum  \\
\hline
3.1          &
2.950--3.350 &
2.900--2.950 &
3.350--3.400 \\
3.57         &
3.560--3.580 &
3.510--3.560 &
3.580--3.630 \\
3.70         &
3.700--3.715 &
3.680--3.700 &
3.715--3.720 \\
3.8          &
3.600--4.000 &
3.500--3.600 &
4.000--4.100 \\
\hline
\end{tabular}
\end{table}

%
% TABLE 6
%
\begin{table}
\caption[]{Equivalent widths (in \AA). The chemical classification into carbon
stars (C) and oxygen-rich stars (o) is based on the presence or absence of the
3.1 $\mu$m feature, supplemented by information from the literature.}
\begin{tabular}{lrrrrc}
\hline\hline
Name             &
W$_{3.1}$        &
W$_{3.57}$       &
W$_{3.70}$       &
W$_{3.8}$        &
C/o              \\
\hline
\multicolumn{6}{l}{\it IR carbon stars} \\
MSX\,LMC\,1042 &
1333           &
$-$0.23        &
5.12           &
405            &
C              \\
MSX\,LMC\,1249 &
889            &
1.79           &
0.04           &
34             &
C              \\
MSX\,LMC\,50   &
786            &
5.04           &
2.16           &
14             &
C              \\
MSX\,LMC\,91   &
1189           &
8.97           &
7.10           &
208            &
C              \\
MSX\,LMC\,93   &
821            &
2.00           &
1.58           &
$-$14          &
C              \\
MSX\,LMC\,55   &
1078           &
6.73           &
2.45           &
221            &
C              \\
MSX\,LMC\,45   &
1236           &
1.72           &
4.79           &
322            &
C              \\
MSX\,LMC\,47   &
1103           &
0.46           &
2.87           &
144            &
C              \\
MSX\,LMC\,219  &
869            &
0.44           &
1.25           &
133            &
C              \\
MSX\,LMC\,232  &
1302           &
0.29           &
1.21           &
182            &
C              \\
MSX\,LMC\,223  &
1016           &
0.96           &
2.76           &
36             &
C              \\
MSX\,LMC\,202  &
1371           &
8.05           &
0.11           &
472            &
C              \\
MSX\,LMC\,225  &
1023           &
0.37           &
2.87           &
101            &
C              \\
MSX\,LMC\,221  &
1783           &
4.38           &
3.11           &
588            &
C              \\
MSX\,LMC\,349  &
877            &
$-$0.62        &
0.02           &
217            &
C              \\
MSX\,LMC\,307  &
1165           &
2.96           &
5.03           &
165            &
C              \\
MSX\,LMC\,341  &
949            &
4.52           &
6.75           &
191            &
C              \\
MSX\,LMC\,484  &
1208           &
6.51           &
3.76           &
140            &
C              \\
MSX\,LMC\,774  &
1144           &
2.80           &
5.03           &
185            &
C              \\
MSX\,LMC\,677  &
1384           &
3.07           &
0.52           &
182            &
C              \\
MSX\,LMC\,635  &
1267           &
7.49           &
3.29           &
400            &
C              \\
MSX\,LMC\,692  &
1023           &
2.44           &
1.40           &
$-$22          &
C              \\
MSX\,LMC\,644  &
1005           &
1.75           &
2.91           &
40             &
C              \\
MSX\,LMC\,654  &
1336           &
8.63           &
6.09           &
157            &
C              \\
MSX\,LMC\,1780 &
1227           &
3.36           &
6.15           &
258            &
C              \\
MSX\,LMC\,743  &
1485           &
0.73           &
3.54           &
377            &
C              \\
MSX\,LMC\,872  &
1046           &
1.55           &
2.35           &
21             &
C              \\
MSX\,LMC\,737  &
1346           &
4.28           &
3.07           &
262            &
C              \\
\hline
\multicolumn{6}{l}{\it IR objects of unknown type} \\
MSX\,LMC\,196 &
493           &
5.20          &
2.35          &
$-$67         &
C\rlap{?}     \\
MSX\,LMC\,733 &
375           &
4.50          &
$-$1.04       &
$-$142        &
o\rlap{?}     \\
\hline
\multicolumn{6}{l}{\it Cluster objects from van Loon et al.\ (2005b)} \\
NGC\,1903-IR1 &
1537          &
6.74          &
5.18          &
448           &
C             \\
NGC\,1903-IR2 &
870           &
3.87          &
1.16          &
70            &
C             \\
NGC\,1984-IR1 &
116           &
0.30          &
$-$0.69       &
$-$29         &
o             \\
NGC\,1984-IR2 &
5             &
$-$0.60       &
$-$0.89       &
248           &
o             \\
NGC\,1978-IR1 &
1110          &
0.70          &
$-$1.89       &
41            &
C             \\
NGC\,1978-IR2 &
1349          &
3.20          &
5.19          &
267           &
C             \\
\hline
\multicolumn{6}{l}{\it Serendipitous objects} \\
MSX\,LMC\,50\,B          &
99                       &
0.94                     &
5.25                     &
37                       &
o                        \\
HV\,11977                &
136                      &
$-$2.04                  &
6.56                     &
72                       &
o                        \\
MSX\,LMC\,349\,B         &
1485                     &
3.50                     &
$-$1.57                  &
$-$23                    &
C                        \\
HV\,2532                 &
$-$38                    &
1.07                     &
1.08                     &
55                       &
o                        \\
MSX\,LMC\,1780\,\rlap{B} &
220                      &
$-$0.75                  &
$-$1.80                  &
198                      &
o                        \\
\hline
\multicolumn{6}{l}{\it IR objects from Matsuura et al.\ (2005)} \\
IRAS\,04286$-$6937 &
508                &
$-$3.66            &
7.14               &
330                &
C                  \\
IRAS\,04496$-$6958 &
1726               &
7.77               &
$-$1.87            &
735                &
C                  \\
IRAS\,04539$-$6821 &
771                &
1.89               &
2.57               &
183                &
C                  \\
IRAS\,04557$-$6753 &
1309               &
0.65               &
3.96               &
542                &
C                  \\
IRAS\,05112$-$6755 &
887                &
0.90               &
5.28               &
224                &
C                  \\
IRAS\,05128$-$6455 &
$-$364             &
1.98               &
$-$1.36            &
$-$228             &
o                  \\
IRAS\,05148$-$6730 &
$-$95              &
$-$5.61            &
$-$1.23            &
286                &
o                  \\
BMB-R\,46          &
1161               &
12.93              &
\llap{$-$}10.33    &
84                 &
C                  \\
LI-LMC\,1813       &
1273               &
16.55              &
$-$4.39            &
151                &
C                  \\
\hline
\end{tabular}
\end{table}

To investigate the band strengths of some of the spectral features, we compute
equivalent widths (Tables 5 \& 6). For the narrow 3.57 and 3.70 $\mu$m bands
it is important to realise that the band definitions are appropriate for our
spectra and hence include the velocity shift of the LMC (0.003--0.004 $\mu$m)
and the internal wavelength calibration accuracy ($\pm$0.002 $\mu$m). In other
data sets these definitions might therefore need a slight readjustment. We
recomputed the equivalent widths at 3.1, 3.57 and 3.8 $\mu$m in the spectra
from Matsuura et al.\ (2005) after converting their F$_\nu$ units to our
F$_\lambda$ units. Different units cause different spectral slopes and hence
different continuum definitions, which affects the equivalent width
measurements. We also made a slight adjustment to the definition of the blue
edge of the 3.1 $\mu$m band.

The uncertainties in the equivalent width values are obtained from the
standard deviations of the values for the eight oxygen-rich objects:
$\sigma=170$, 2.23, 3.03 and 175 \AA\ in the equivalent widths at 3.1, 3.57,
3.70 and 3.8 $\mu$m, respectively. This assumes that oxygen-rich objects have
zero equivalent widths, which is not strictly true; the quoted uncertainties
are therefore conservative estimates. The 3.57 and especially the 3.70 $\mu$m
features are quite difficult to measure in individual spectra.

All four groups of IR carbon stars show a very clear correlation between the
strength of the 3.8 $\mu$m band and that of the 3.1 $\mu$m band (Fig.\ 12),
with a Pearson correlation coefficient of $p=0.80$: an increase in column
density of C$_2$H$_2$ would make both bands appear stronger. The 3.1 $\mu$m
band is always prominent, within a range in equivalent width of a factor two.
The 3.8 $\mu$m band is always seen in conjunction with strong 3.1 $\mu$m
absorption, $W_{3.1}>1000$ \AA\ ($p=0.85$, including the additional stars ---
crosses in Fig.\ 12), but it is not always visible in spectra with weaker 3.1
$\mu$m absorption, when the 3.1 $\mu$m band may be dominated by the
contribution from HCN.

For $W_{3.8}<450$ \AA, the correlation between the 3.70 and 3.8 $\mu$m band
strengths is much better ($p=0.41$) than that between the 3.57 and 3.8 $\mu$m
band strengths ($p=0.11$). This might imply that the 3.70 $\mu$m band is
associated with C$_2$H$_2$. At $W_{3.8}>450$ \AA\ veiling and changes in
excitation conditions (see below) may become important and affect the
correlations.

%
% FIGURE 12
%
\begin{figure*}[tb]
\centerline{\hbox{
\psfig{figure=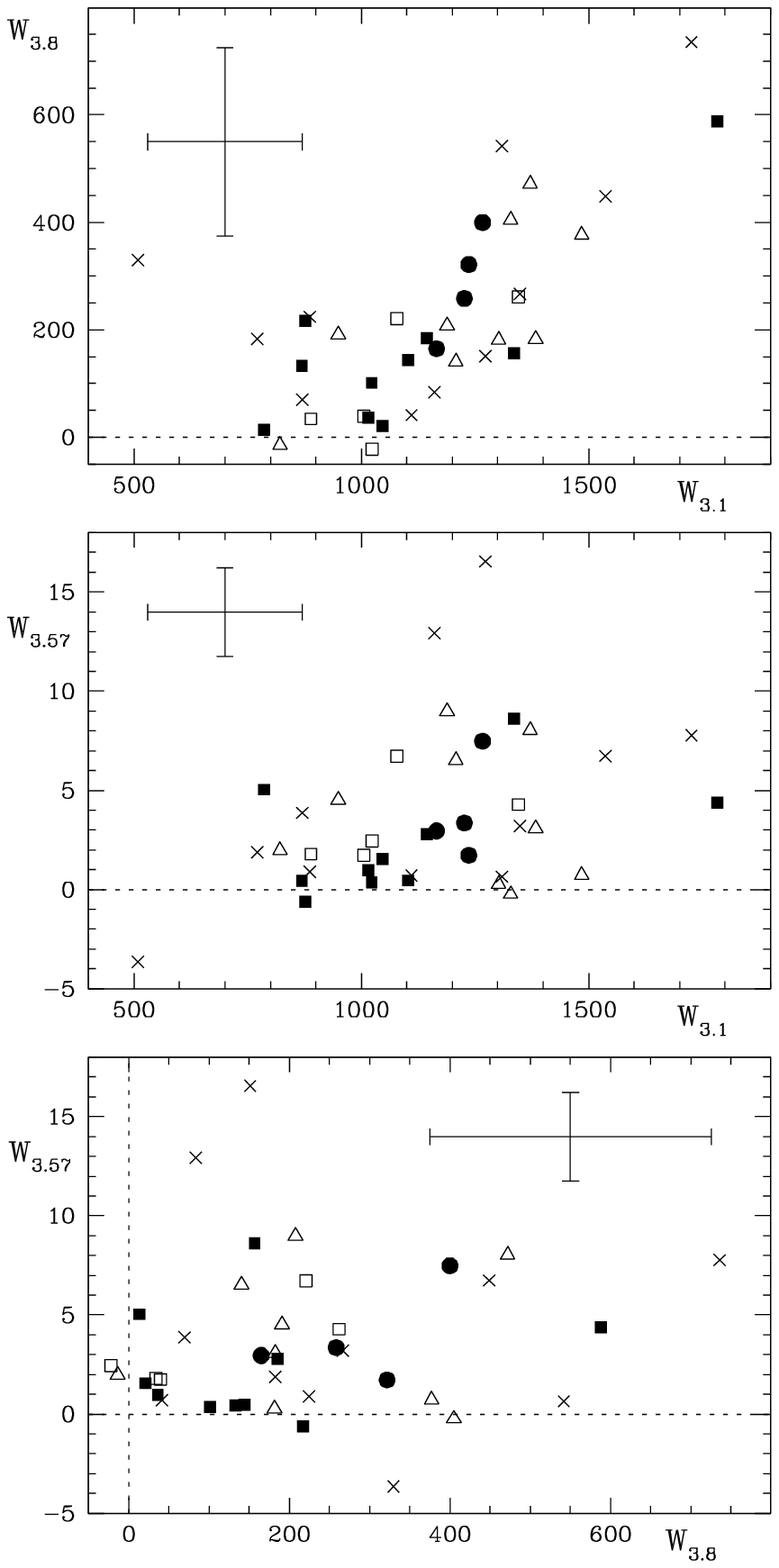,width=80mm}\hspace{10mm}
\psfig{figure=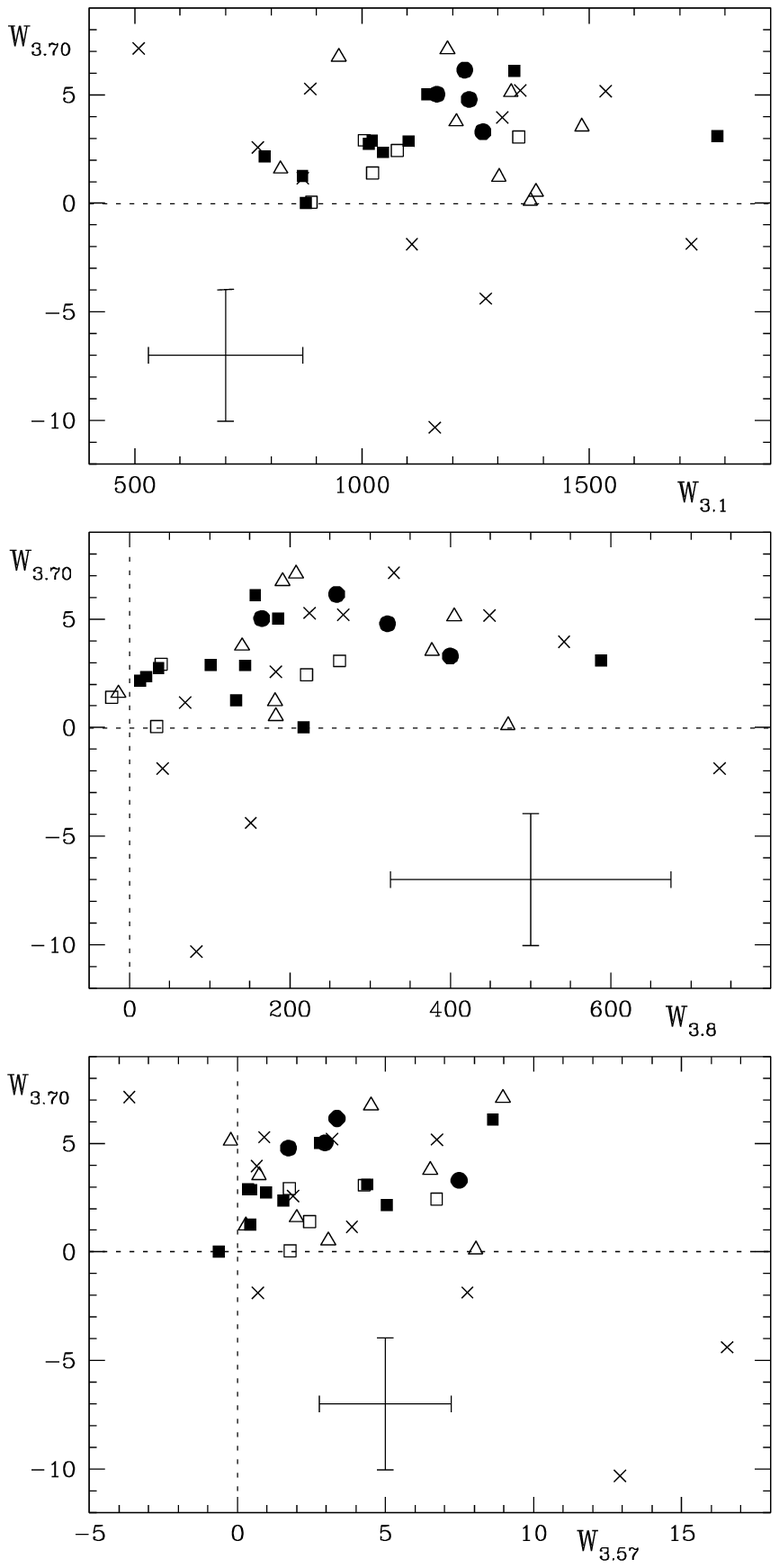,width=80mm}
}}
\caption[]{Equivalent widths of the 3.1, 3.57, 3.70 and 3.8 $\mu$m bands (in
\AA) plotted versus each other. A typical errorbar is plotted in each panel.
The strongest correlation is between the 3.1 and 3.8 $\mu$m bands (top left).
Symbols are as in Fig.\ 7.}
\end{figure*}

Some of the scatter in the equivalent width distributions is intrinsic. For
instance, MSX\,LMC\,91 exhibits extremely strong 3.57 and 3.70 $\mu$m bands
but only moderately strong 3.1 and 3.8 $\mu$m bands, whilst MSX\,LMC\,221
shows exactly the opposite behaviour. Another extreme object is the relatively
blue MSX\,LMC\,1249, whose 3.1 $\mu$m band resembles that of water ice (which
is not expected to exist in a warm carbon-dominated chemistry) and which does
not exhibit any 3.70 or 3.8 $\mu$m absorption although it does show a small
dip at 3.57 $\mu$m. The differences between extreme examples of 3--4 $\mu$m
spectra (Fig.\ 10) are much greater than the subtle differences as a function
of luminosity and mass-loss rate (Fig.\ 11), suggesting that other parameters
influence the shape and strength of the molecular bands.

The shape of the spectrum between about 3.1 and 3.7 $\mu$m can be dramatically
different for different carbon stars (see Fig.\ 10), without it being clear
what absorption band --- if any --- is responsible for this. To investigate
this further we compute the spectral slopes, $s(\lambda)$, around 3.15, 3.24,
3.43 and 3.57 $\mu$m within the regions 3.12--3.18, 3.20--3.28, 3.36--3.50 and
3.50--3.64 (excluding 3.56--3.58) $\mu$m, respectively. We then compute the
pair-wise differential spectral slopes, $s^\prime(\lambda)$ (Table 7), to
quantify the degree to which the spectrum displays a ``knee'' such as is seen,
for example, at 3.2 $\mu$m in MSX\,LMC\,307 or at 3.5 $\mu$m in MSX\,LMC\,221.
If there is no change in spectral slope the differential spectral slope is
zero.

%
% TABLE 7
%
\begin{table}
\caption[]{Differential spectral slopes $s^\prime(3.2)=s(3.15)-s(3.24)$ and
$s^\prime(3.5)=s(3.43)-s(3.57)$ (see text), and the fractional contributions
of the dust emission, for carbon stars.}
\begin{tabular}{lccccc}
\hline\hline
Name            &
s$^\prime$(3.2) &
s$^\prime$(3.5) &
f$_{3.0}$       &
f$_{3.5}$       &
f$_{4.0}$       \\
\hline
\multicolumn{6}{l}{\it IR carbon stars} \\
MSX\,LMC\,1042 &
2.59           &
0.61           &
0.36           &
0.56           &
0.70           \\
MSX\,LMC\,1249 &
1.65           &
0.28           &
0.48           &
0.64           &
0.74           \\
MSX\,LMC\,50   &
2.42           &
0.38           &
0.47           &
0.67           &
0.78           \\
MSX\,LMC\,91   &
1.73           &
1.01           &
0.45           &
0.64           &
0.76           \\
MSX\,LMC\,93   &
1.73           &
0.44           &
0.52           &
0.68           &
0.78           \\
MSX\,LMC\,55   &
2.32           &
0.57           &
0.80           &
0.86           &
0.89           \\
MSX\,LMC\,45   &
1.50           &
0.96           &
0.47           &
0.66           &
0.78           \\
MSX\,LMC\,47   &
1.68           &
0.28           &
0.46           &
0.65           &
0.77           \\
MSX\,LMC\,219  &
1.29           &
0.82           &
0.54           &
0.70           &
0.80           \\
MSX\,LMC\,232  &
1.77           &
0.24           &
0.37           &
0.57           &
0.71           \\
MSX\,LMC\,223  &
3.78           &
0.31           &
0.55           &
0.71           &
0.81           \\
MSX\,LMC\,202  &
0.23           &
1.71           &
0.54           &
0.70           &
0.80           \\
MSX\,LMC\,225  &
1.88           &
0.36           &
0.47           &
0.67           &
0.78           \\
MSX\,LMC\,221  &
1.10           &
1.49           &
0.37           &
0.57           &
0.71           \\
MSX\,LMC\,349  &
1.40           &
0.66           &
0.56           &
0.73           &
0.82           \\
MSX\,LMC\,307  &
5.14           &
0.07           &
0.47           &
0.66           &
0.78           \\
MSX\,LMC\,341  &
2.68           &
1.06           &
0.54           &
0.70           &
0.80           \\
MSX\,LMC\,484  &
2.16           &
0.83           &
0.85           &
0.89           &
0.92           \\
MSX\,LMC\,774  &
2.26           &
0.64           &
0.37           &
0.58           &
0.72           \\
MSX\,LMC\,677  &
2.72           &
1.05           &
0.51           &
0.67           &
0.78           \\
MSX\,LMC\,635  &
1.03           &
1.60           &
0.47           &
0.66           &
0.78           \\
MSX\,LMC\,692  &
1.84           &
0.53           &
0.52           &
0.68           &
0.78           \\
MSX\,LMC\,644  &
3.01           &
0.38           &
0.52           &
0.68           &
0.78           \\
MSX\,LMC\,654  &
4.50           &
0.46           &
0.57           &
0.73           &
0.82           \\
MSX\,LMC\,1780 &
3.30           &
0.21           &
0.38           &
0.60           &
0.75           \\
MSX\,LMC\,743  &
1.55           &
1.23           &
0.53           &
0.69           &
0.79           \\
MSX\,LMC\,872  &
1.36           &
0.53           &
0.54           &
0.71           &
0.81           \\
MSX\,LMC\,737  &
0.47           &
0.82           &
0.51           &
0.67           &
0.77           \\
\hline
\multicolumn{6}{l}{\it Cluster carbon stars from van Loon et al.\ (2005b)} \\
NGC\,1903-IR1  &
0.60           &
1.04           &
0.79           &
0.83           &
0.85           \\
NGC\,1903-IR2  &
\llap{$-$}1.76 &
0.53           &
0.74           &
0.77           &
0.79           \\
NGC\,1978-IR1  &
2.09           &
0.78           &
0.63           &
0.71           &
0.76           \\
NGC\,1978-IR2  &
2.98           &
0.97           &
0.76           &
0.82           &
0.86           \\
\hline
\multicolumn{6}{l}{\it IR carbon-rich stars from Matsuura et al.\ (2005)} \\
IRAS\,04286$-$6937 &
5.18               &
0.10               &
0.06               &
0.19               &
0.38               \\
IRAS\,04496$-$6958 &
1.57               &
1.08               &
0.48               &
0.62               &
0.72               \\
IRAS\,04539$-$6821 &
2.25               &
0.05               &
0.51               &
0.68               &
0.78               \\
IRAS\,04557$-$6753 &
3.43               &
0.76               &
0.44               &
0.65               &
0.78               \\
IRAS\,05112$-$6755 &
1.38               &
0.18               &
0.18               &
0.38               &
0.57               \\
BMB-R\,46          &
0.06               &
1.20               &
0.16               &
0.25               &
0.34               \\
LI-LMC\,1813       &
2.06               &
0.72               &
0.38               &
0.59               &
0.74               \\
\hline
\end{tabular}
\end{table}

%
% FIGURE 13
%
\begin{figure}[tb]
\centerline{\psfig{figure=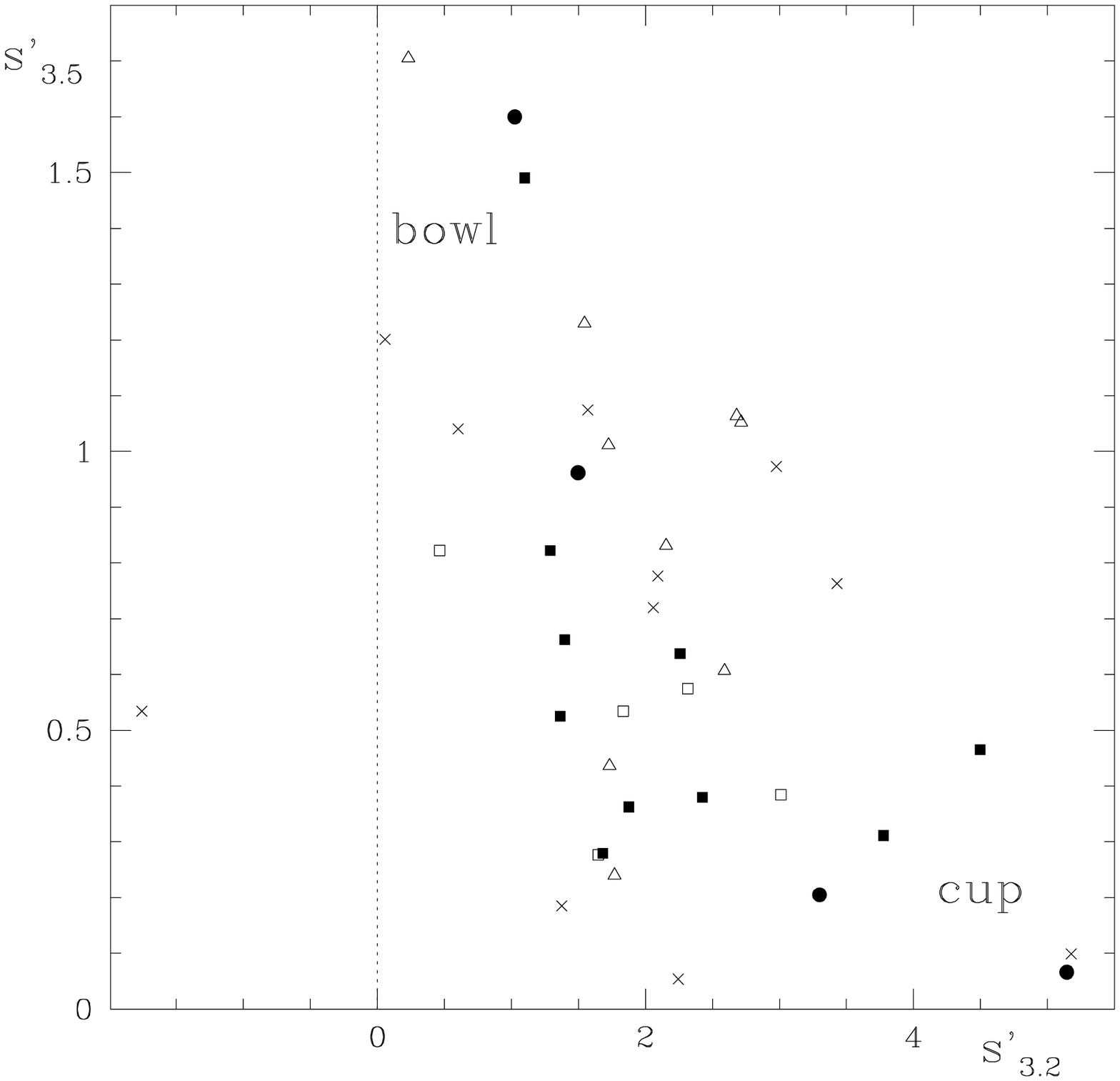,width=88mm}}
\caption[]{Change in slope around 3.5 $\mu$m versus that around 3.2 $\mu$m;
there is a clear correlation. Symbols are as in Fig.\ 7.}
\end{figure}

The differential slopes at 3.2 and 3.5 $\mu$m show a strong anti-correlation
(Fig.\ 13; $p=-0.57$, or $p=-0.62$ if the noisier spectra of group 3 sources
are omitted). There is a gradual transition between two extreme types of
spectral shape: those with a narrow 3.1 $\mu$m band and a flat continuum
between 3.2 and 3.7 $\mu$m (``cup''), and those with absorption extending from
the 3.1 $\mu$m band all the way until 3.5 $\mu$m where it meets the continuum
or dips into the 3.8 $\mu$m band (``bowl''). Interestingly, the best examples
of the ``cup'' type are all moderate to high mass-loss rate and luminosity
objects (including IRAS\,04286$-$6937) although the ``bowl'' type also occurs
amongst these classes of stars.

%
% FIGURE 14
%
\begin{figure}[tb]
\centerline{\psfig{figure=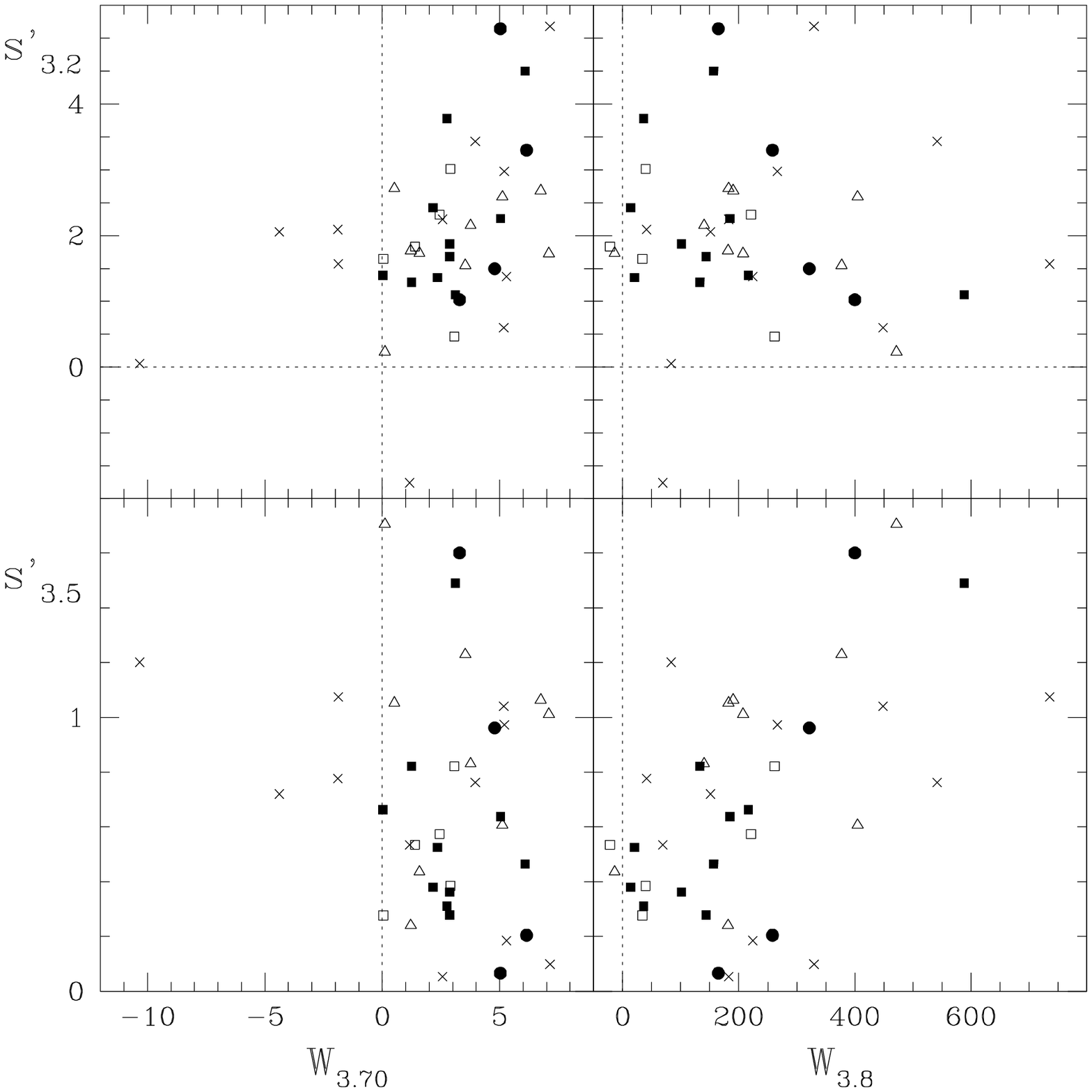,width=88mm}}
\caption[]{Change in slope around 3.2 and 3.5 $\mu$m versus the equivalent
width of the 3.70 and 3.8 $\mu$m bands. The 3.70 $\mu$m band is generally
stronger for larger $s^\prime_{3.2}$ and smaller $s^\prime_{3.5}$, whilst the
3.80 $\mu$m band shows the opposite behaviour. Symbols are as in Fig.\ 7.}
\end{figure}

The strength of the 3.70 $\mu$m feature is correlated with the differential
slope at 3.2 $\mu$m (Fig.\ 14; $p=0.45$). This suggests that whatever causes
the broad absorption between 3.2 and 3.5 $\mu$m is anti-correlated with the
3.70 $\mu$m feature. On the other hand the strength of the 3.8 $\mu$m band is
correlated with the differential slope at 3.5 $\mu$m ($p=0.72$) and mildly
anti-correlated with that at 3.2 $\mu$m ($p=-0.33$), suggesting a link between
the 3.2--3.5 and 3.8 $\mu$m absorptions. We interpret this as an effect of
different excitation conditions. A bowl-shaped 3.1 $\mu$m band indicates a
high excitation temperature, which would then also result in broad absorption
by C$_2$H$_2$ at 3.8 $\mu$m. A low excitation temperature would result in
narrow absorption features of C$_2$H$_2$, between 3.70--3.78 $\mu$m.

%------------------------------------------------------------------------- 4.4
\subsection{Relation between the molecular atmosphere and circumstellar dust
envelope}

Can some of the differences in the molecular bands be reflected in differences
in the circumstellar dust envelope? Stars with the highest mass-loss rates and
luminosities always have strong 3.8 $\mu$m bands and almost always do they
also have a conspicuous 3.70 $\mu$m feature (Fig.\ 15). Stars with low
mass-loss rates of $\dot{M}<10^{-5}$ M$_\odot$ yr$^{-1}$ tend to have weak or
absent 3.70 and 3.8 $\mu$m absorptions whilst the 3.57 $\mu$m band is often
visible in these stars.

%
% FIGURE 15
%
\begin{figure}[tb]
\centerline{\psfig{figure=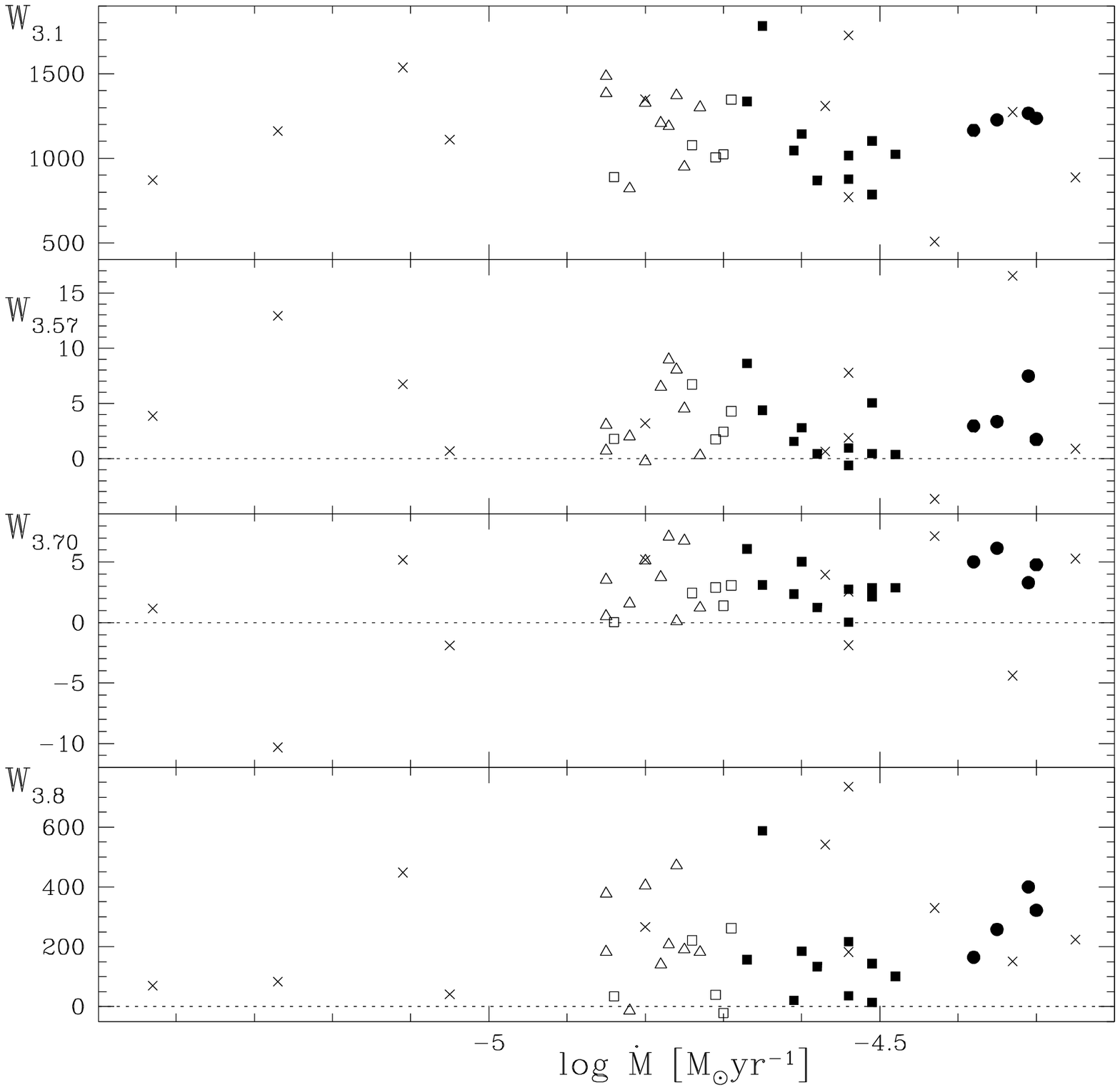,width=88mm}}
\caption[]{Equivalent widths versus mass-loss rate. There is no clear
correlation other than that the absorption is generally strong at the highest
mass-loss rates. Symbols are as in Fig.\ 7.}
\end{figure}

At high mass-loss rates of $\dot{M}>2.5\times10^{-5}$ M$_\odot$ yr$^{-1}$ the
3.1, 3.57 and 3.8 $\mu$m absorption bands increase in strength with increasing
mass-loss rate. However, at lower mass-loss rates the bands are generally
stronger than around $3\times10^{-5}$ M$_\odot$ yr$^{-1}$, albeit with
considerable spread. The same trends are seen for the 3.70 $\mu$m band. This
probably reflects a competition between the effect of veiling (reducing the
equivalent width) and an optically thicker atmosphere in connection with a
higher mass-loss rate (increasing the equivalent width).

%
% FIGURE 16
%
\begin{figure}[tb]
\centerline{\psfig{figure=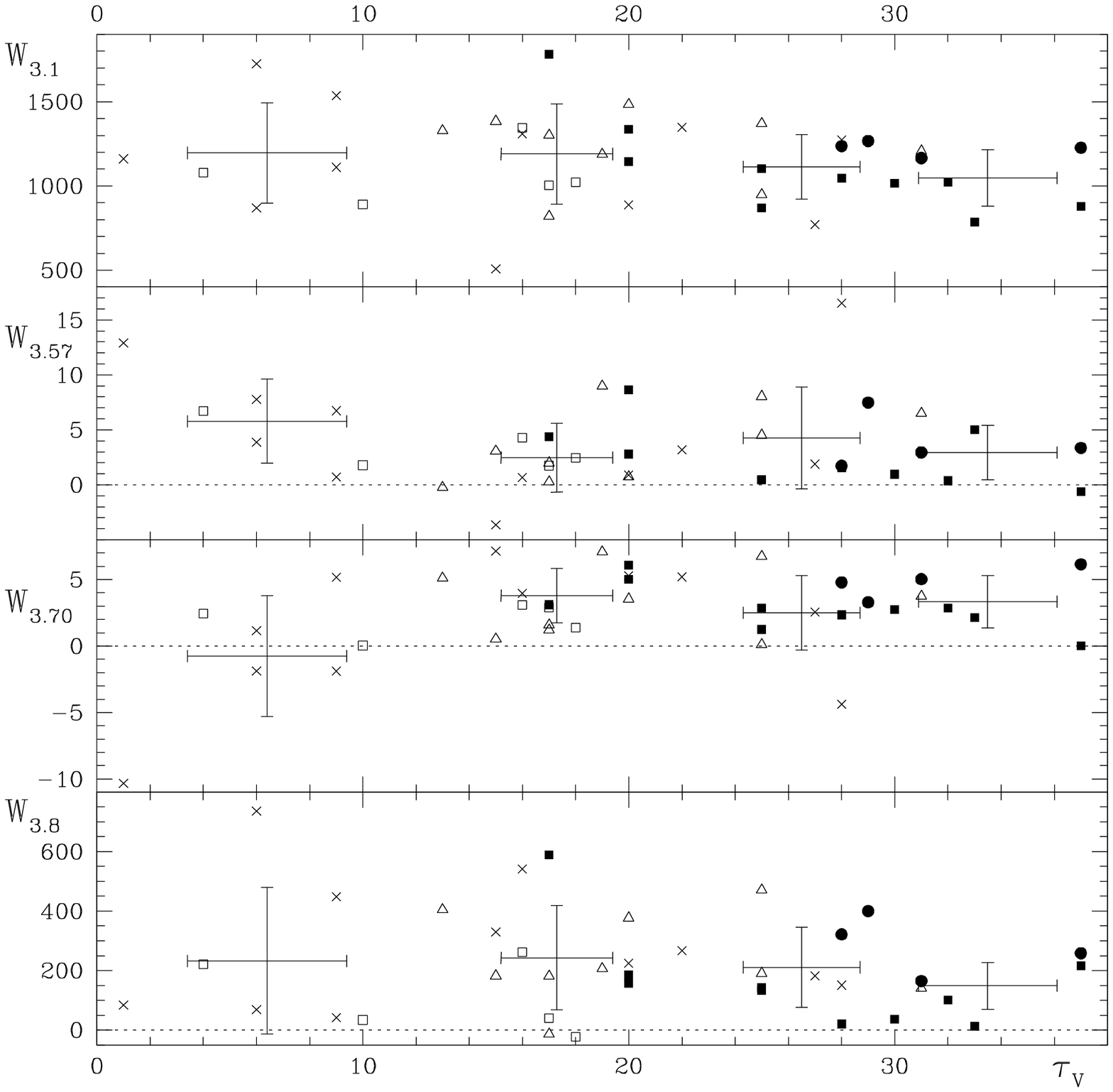,width=88mm}}
\caption[]{Equivalent widths versus dust optical depth. The error bars refer
to the mean values and standard deviations in the intervals $\tau_{\rm
V}\in\langle0,10]$, $\langle10,20]$, $\langle20,30]$ and
$\langle30,\rightarrow\rangle$. There are weak trends in the 3.1 and 3.8
$\mu$m bands to diminish in strength with greater optical depth. Symbols are
as in Fig.\ 7.}
\end{figure}

Dust column densities increase with increasing mass-loss rate but also with
decreasing luminosity as low-luminosity objects are more compact. Molecular
column densities may be affected in a similar way, and the dust optical depth
might therefore show a stronger correlation with the molecular column density
than either the mass-loss rate or luminosity would do. The 3.1 and 3.8 $\mu$m
bands become weaker at progressively larger dust optical depth (Fig.\ 16).
This is likely due to veiling by the increasing contribution of dust emission.
The decrease in equivalent width is indeed larger at 3.8 $\mu$m than at 3.1
$\mu$m. The 3.57 and 3.70 $\mu$m equivalent width seems to remain constant,
though, which may suggest that they would be seen to increase with optical
depth if the effects of veiling were to be removed. At relatively modest
optical depth, $\tau_{\rm V}<10$, the 3.57 $\mu$m band is often quite strong
but the 3.70 $\mu$m band is often absent.

%
% FIGURE 17
%
\begin{figure}[tb]
\centerline{\psfig{figure=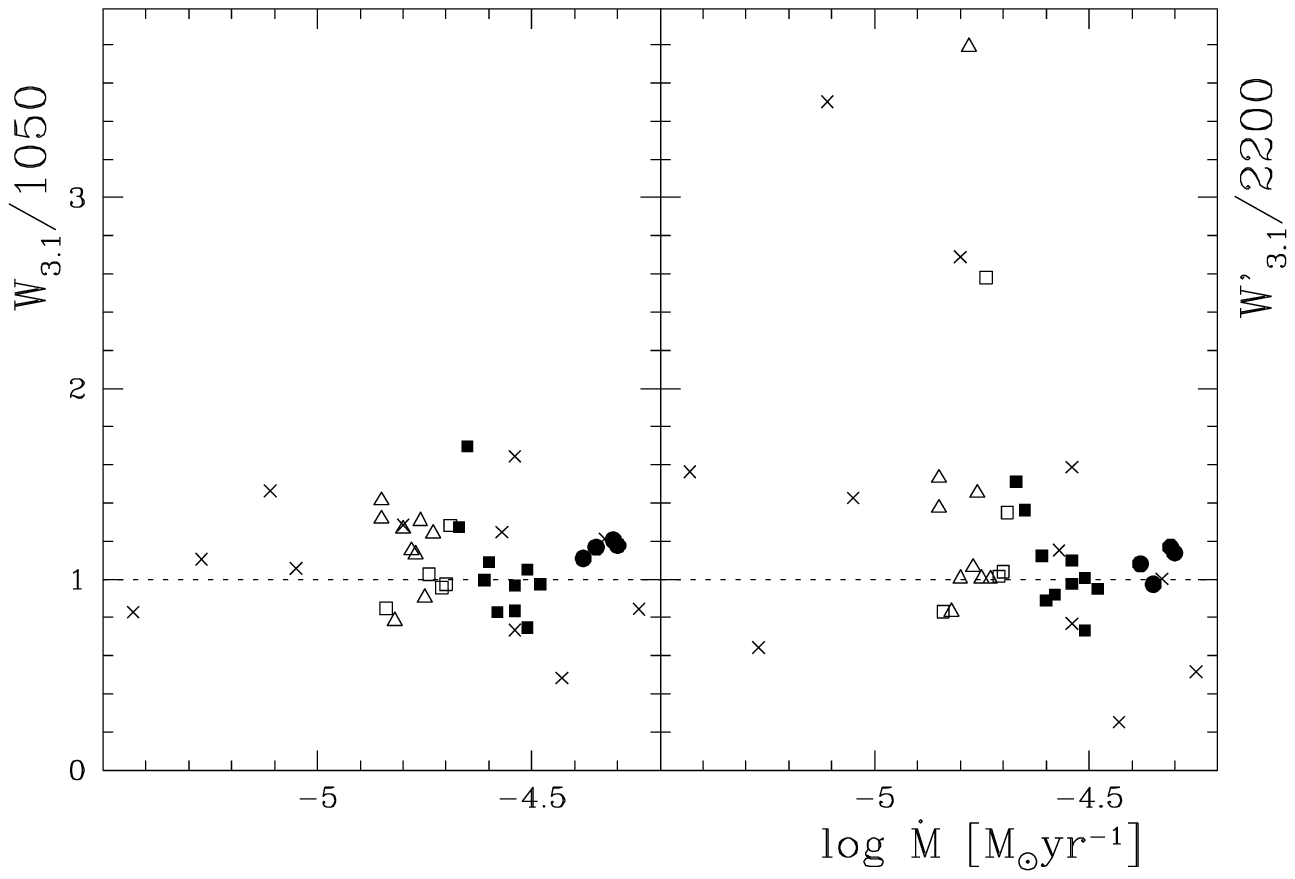,width=88mm}}
\caption[]{Equivalent width of the 3.1 $\mu$m band versus mass-loss rate
before (left) and after (right) correction for dust emission, normalised to
the peak in the distribution. Removing the dust emission creates a flatter and
tighter distribution around unity for about half of the sample, but it creates
large scatter for the remainder --- many of which have low mass-loss rates.
Symbols are as in Fig.\ 7.}
\end{figure}

The effects of veiling by dust emission can in principle be removed, as {\sc
dusty} returns the fractional contribution of the dust emission to the total
flux density as a function of wavelength. We tabulate these values at
wavelengths of 3.0, 3.5 and 4.0 $\mu$m (Table 7). Unfortunately, these
fractions are ill-determined: they depend sensitively on the amount and
temperature of the warmest dust, which is also the most time-variable
component. Hence correcting the equivalent widths in this way increases the
overall scatter in a sub-set of the sample of carbon stars (Fig.\ 17), which
is especially due to stars with fairly modest mass-loss rates. However, for at
least half the sample of stars the scatter is reduced and the 3.1 $\mu$m
equivalent widths of the four groups of stars are more in line with each other
(Fig.\ 17). This suggests that veiling is the main driver behind variations of
the 3.1 $\mu$m equivalent width with mass-loss rate and luminosity.

Consequently the photospheric 3.1 $\mu$m equivalent width seems to depend very
little on stellar or circumstellar parameters, and we propose that this band
may be saturated for the MSX-selected sample of IR carbon stars. Comparing the
dust continuum contribution (Table 7) with the maximum depth of the 3.1 $\mu$m
absorption in the spectra of Fig.\ 2, it appears that part of the absorption
may in fact be seen against the dust continuum. If true this would imply a
dense molecular atmosphere with a scale height of order a few stellar radii,
which is seen in absorption against the dust emission arising from behind and
adjacent to the limb of the star. The same may be true for the 3.8 $\mu$m band
in some stars.

A large molecular zone is also implied by the low temperature at the dust
formation radius that was required to fit the SEDs. The molecular atmosphere
must extend to at least the dust formation radius for dust to form, and a
lower temperature at that radius implies that the dust forms farther from the
star. If the temperature had been underestimated, then the veiling of the
absorption bands by the dust continuum around 3--4 $\mu$m would also have been
underestimated. This would make the case even more convincing for part of the
molecular absorption to be seen against background dust continuum emission.

%
% FIGURE 18
%
\begin{figure}[tb]
\centerline{\psfig{figure=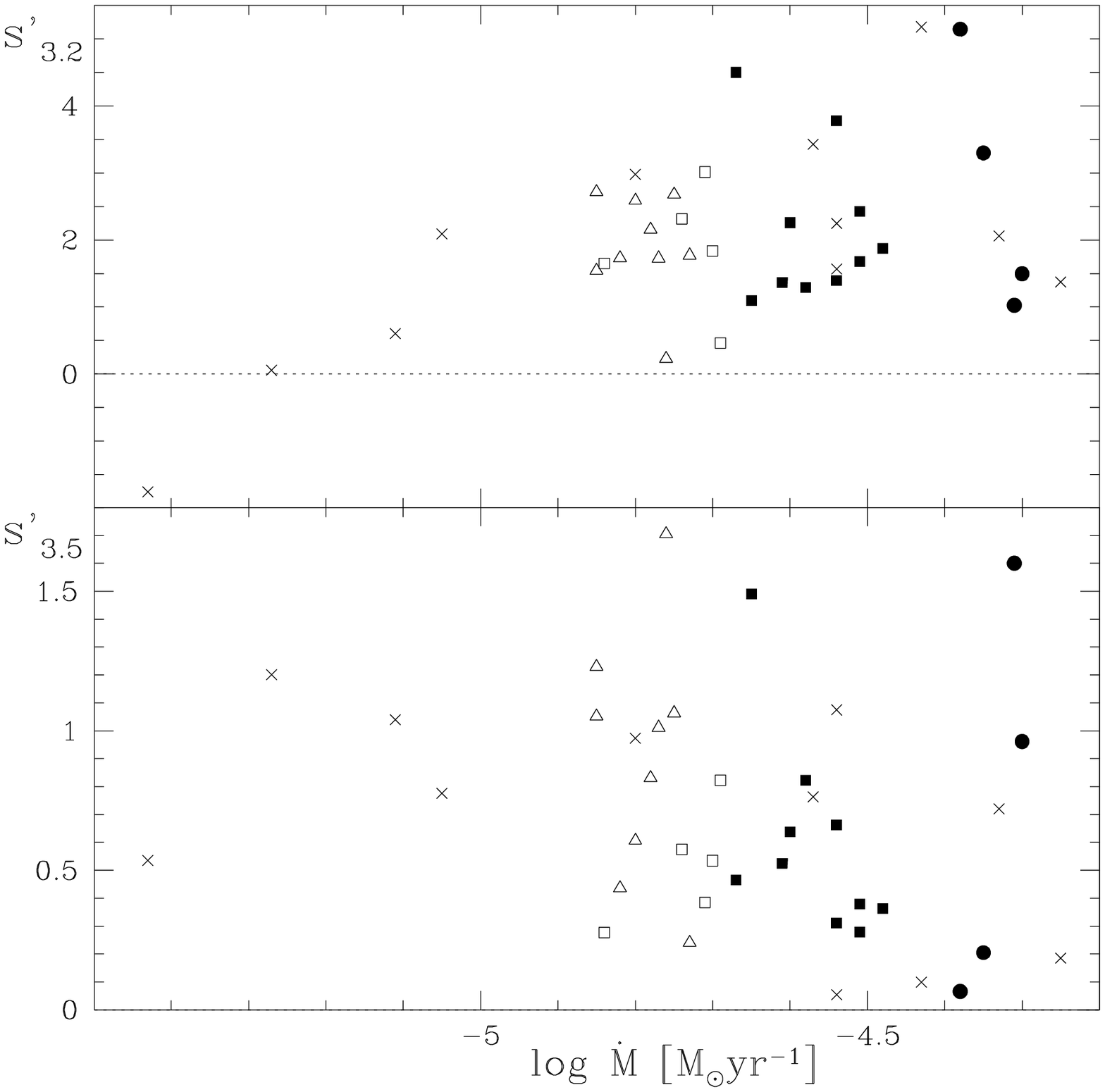,width=88mm}}
\caption[]{Change in slope around 3.2 and 3.5 $\mu$m versus mass-loss rate.
There is a trend for the spectral shape in the 3.1--3.7 $\mu$m region to turn
from ``bowl'' to ``cup'', although the two MSX-selected carbon stars with the
highest mass-loss rates exhibit a ``bowl'' shaped spectrum. Symbols are as in
Fig.\ 7.}
\end{figure}

A weak trend can be seen in the differential spectral slope at 3.5 $\mu$m as a
function of mass-loss rate (Fig.\ 18), suggesting that stars enduring higher
mass-loss rates have a ``cup''-shaped 3.2--3.5 $\mu$m spectrum. At high
mass-loss rate the 3.1 $\mu$m band may on average be formed higher above the
stellar photosphere, at a lower excitation temperature, which would explain
the narrower shape (``cup''). On the other hand, the two MSX-selected stars
with the highest mass-loss rates defy such relationship.

%------------------------------------------------------------------------- 4.5
\subsection{Pulsation and excitation conditions}

Pulsation is known to affect the absorption bands. The difficulty with
investigating dependencies on pulsational properties is that the effects have
a highly time-variable component. But the pulsation may also affect
time-averaged properties, that we can study here. In addition to the stars
with known pulsation properties listed in Table 2, pulsation periods and
amplitudes are known for IRAS\,04496$-$6958 and IRAS\,04557$-$6753 (Whitelock
et al.\ 2003) and rough estimates for LI-LMC\,1813 (van Loon et al.\ 2003).
The L$^\prime$-band amplitude of 1.6 mag for LI-LMC\,1813 is scaled to a
K$_{\rm s}$-band amplitude of ${\Delta}K_{\rm s}\sim2.0$ mag (cf.\ Whitelock
et al.\ 2003).

%
% FIGURE 19
%
\begin{figure}[tb]
\centerline{\psfig{figure=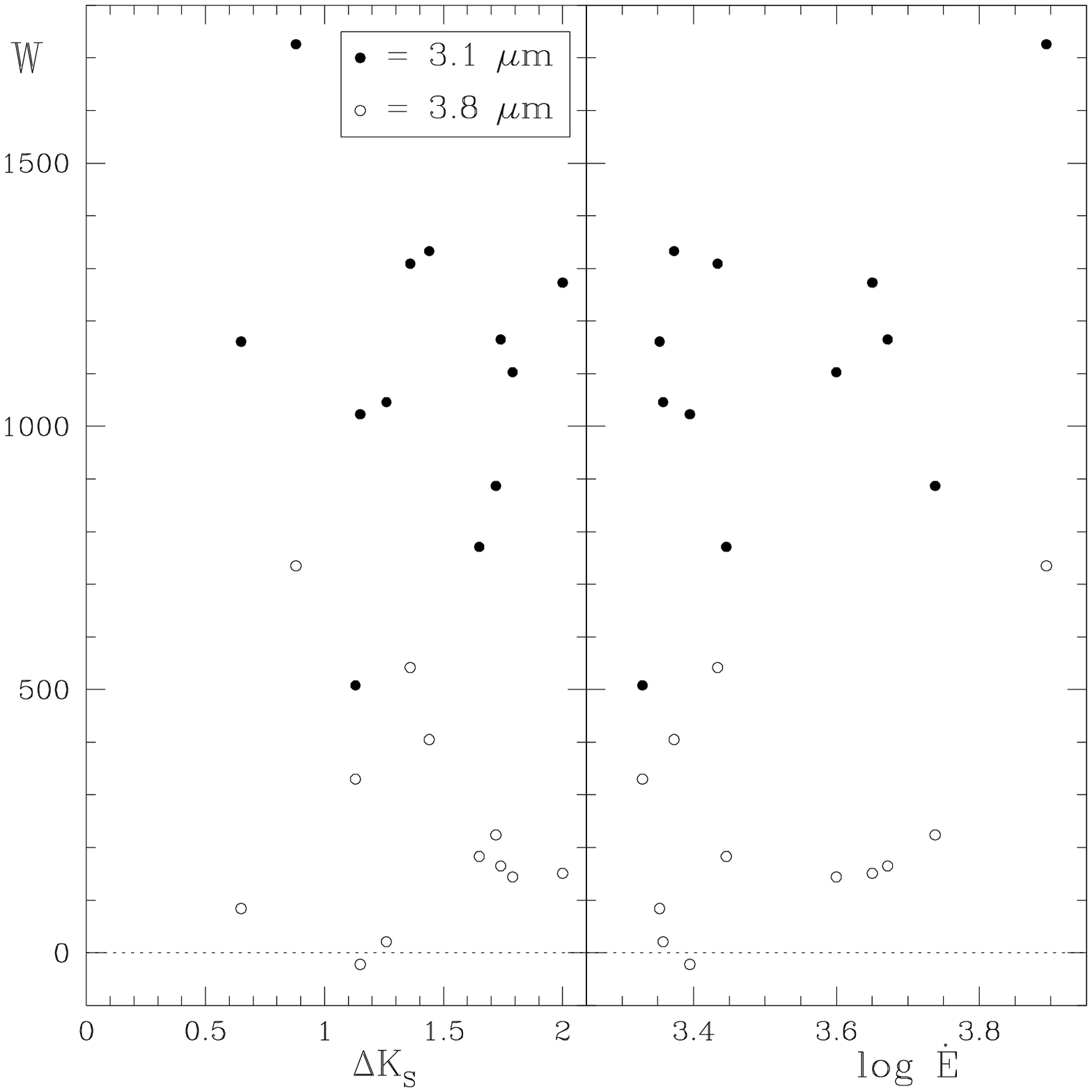,width=88mm}}
\caption[]{Equivalent widths at 3.1 $\mu$m (solid symbols) and 3.8 $\mu$m
(open symbols) versus K$_{\rm s}$-band pulsation amplitude (left panel), and
versus pulsational energy injection rate per cycle, $\dot{E}$ (right panel),
in logarithmic units of L$_\odot$. Whilst no correlation is seen with
amplitude, the equivalent width is larger for larger energy injection rates.}
\end{figure}

The equivalent width (at 3.1 and 3.8 $\mu$m) shows no correlation with the
photometric amplitude, ${\Delta}K_{\rm s}$ (Fig.\ 19, left panel; $p=-0.15$
and $p=-0.21$, respectively). However, the photometric amplitude is relative
to the star's luminosity. An absolute quantity that more directly measures the
amount of energy involved in the pulsation is the energy injection rate (van
Loon 2002):
\begin{equation}
\dot{E} = \frac{L}{2} \times \left\{ \frac{\exp \left( {\Delta}K_{\rm s}/2.5
\right) - 1}{\exp \left( {\Delta}K_{\rm s}/2.5 \right) + 1} \right\}
\end{equation}
This is the fraction of the bolometric luminosity that is used in expanding
the star, assuming that the amplitude at K$_{\rm s}$ measures the bolometric
amplitude and that the luminosity varies sinusoidally. It is by definition
limited to $0<\dot{E}<L/2$. The equivalent widths at 3.1 and 3.8 $\mu$m show a
weak correlation with $\dot{E}$ (Fig.\ 19, right panel; $p=0.50$ and $p=0.41$,
respectively), suggesting a positive link between stronger pulsation and a
larger column density of the molecular atmosphere --- in particular of
C$_2$H$_2$.

%
% FIGURE 20
%
\begin{figure}[tb]
\centerline{\psfig{figure=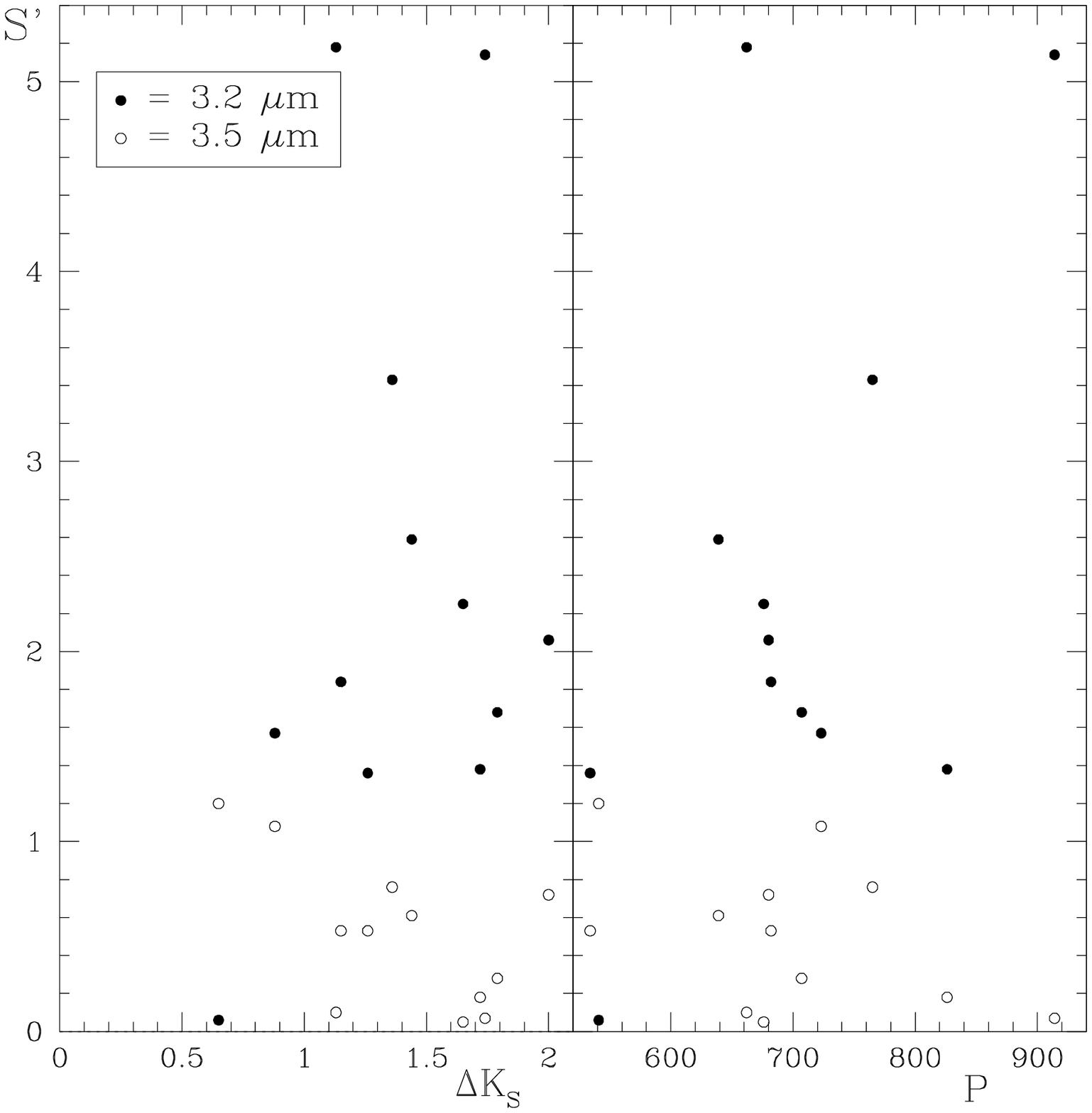,width=88mm}}
\caption[]{Change in slope at 3.2 $\mu$m (solid symbols) and 3.5 $\mu$m (open
symbols) versus K$_{\rm s}$-band pulsation amplitude (left panel), and versus
pulsation period (right panel). The 3.2--3.5 $\mu$m absorption wing is weaker
for larger amplitudes and longer periods, indicating a lower excitation
temperature in a more extended molecular atmosphere.}
\end{figure}

Pulsation also affects the excitation conditions. The fact that the 3.1 $\mu$m
band is narrower at high mass-loss rates is because both are related to the
increased scale height as a consequence of strong pulsation. Indeed (Fig.\
20), the differential spectral slope at 3.2 $\mu$m is weakly correlated with
the K$_{\rm s}$-band amplitude ($p=0.26$) and pulsation period ($p=0.53$)
whilst the differential spectral slope at 3.5 $\mu$m shows the opposite
behaviour ($p=-0.61$ and $p=-0.45$, respectively). This suggests that the 3.1
$\mu$m absorption profile is narrower at larger photometric amplitude and
longer pulsation period, supporting evidence for a lower excitation
temperature in a more extended molecular atmosphere. There is a possible
competing mechanism associated with the pulsation: shocks induced by strong
pulsation cause the band to broaden again (Yamamura et al.\ 1997). This might
explain the ``bowl''-shaped absorption in the two MSX-selected stars with the
highest mass-loss rates.

%=========================================================================== 5
\section{Discussion}

\subsection{The population of dust-enshrouded carbon stars}

The population of visually bright carbon stars in the LMC, selected on the
basis of objective prism surveys, is estimated to be $N_{\rm vis}\simeq9,000$
(Kontizas et al.\ 2001). In the galaxy, many carbon stars become extremely
obscured by the very opaque carbon dust they produce when they enter the
superwind stage with $\dot{M}>10^{-5}$ M$_\odot$ yr$^{-1}$ (e.g., Jura \&
Kleinmann 1990). These objects would escape detection in the optical and
(shallow, wide-field) near-IR surveys in the LMC but could stand out at mid-IR
wavelengths where their dust envelopes shine brightly. The IRAS Point Source
Catalogue was searched for obscured AGB stars in the LMC, and although several
hundred such candidates were found (Loup et al.\ 1997) only a few dozen were
successfully identified with carbon stars, with another few dozen oxygen-rich
massive AGB stars or red supergiants (Reid et al.\ 1990; Wood et al.\ 1992;
Zijlstra et al.\ 1996; van Loon et al.\ 1997). This was but a small fraction
of the expected population of superwind AGB stars. However, the relatively
deep near-IR imaging of IRAS positions by van Loon et al.\ (1997) recovered
counterparts all the way down to their detection limit of $K_{\rm s}<18$ mag,
whilst leaving half of their IRAS sample unidentified. Most of the faint
counterparts had colours of carbon stars --- later spectroscopic data
confirmed their carbon star nature (van Loon et al.\ 1999a; Trams et al.\
1999). Van Loon et al.\ (1997) took this as evidence for a hitherto uncovered
population of very obscured carbon stars.

The MSX Point Source Catalogue at 8.3 $\mu$m was cross-identified with the
2MASS JHK$_{\rm s}$ catalogue to reveal the largest selection to date of
candidate superwind AGB stars in the LMC (Egan et al.\ 2001; Wood \& Cohen
2001). Several hundred were found, with more than a hundred remaining without
near-IR counterpart. Given the result of van Loon et al.\ (1997) and the
shallower depth of 2MASS at $K_{\rm s}<15$ mag we can expect many of the
MSX-selected candidate AGB stars and unidentified stellar MSX sources to be
dust-enshrouded (IR) carbon stars. We here provide supporting evidence for
this hypothesis, as our selection of targets from the MSX-2MASS compilation
without optical counterparts turns out to almost entirely consist of IR carbon
stars (at least 28 out of 30).

Allowing for incompleteness in the MSX catalogue due to the detection
threshold and confusion in crowded areas or areas with a complex background,
the total population of IR carbon stars in the LMC may approach $N_{\rm
IR}\simeq1,000$. This would suggest that, typically, a 1.5--3 M$_\odot$ star
spends around 10 per cent of its thermal-pulsing AGB lifetime being heavily
dust-enshrouded. With mass-loss rates in this superwind stage enhanced by a
factor of 10 (or more) over those experienced during the unobscured stages,
the mass return during the superwind stage of a carbon star must be (at least)
50 per cent of the total mass returned over its lifetime. This is in excellent
agreement with the results found for cluster IR objects (van Loon et al.\
2005b).

\subsection{Spectral diagnostics}

The 3.1 $\mu$m band is due to a mixture of HCN and C$_2$H$_2$, but it is
difficult to determine the relative contributions of these molecules. In all
MSX-selected IR carbon stars in the LMC the 3.8 $\mu$m band peaks at 3.75
$\mu$m and is therefore probably due to C$_2$H$_2$; it never peaks at 3.9
$\mu$m as it often does in the galaxy due to absorption by HCN and/or CS. The
lower metallicity in the LMC would yield lower nitrogen and sulphur
abundances, reducing the potential for forming HCN or CS. The equivalent
widths of the 3.1 and 3.8 $\mu$m bands correlate very well, although the 3.8
$\mu$m band is not always seen whilst the 3.1 $\mu$m is always prominent. This
suggests that when both bands are strong, the C$_2$H$_2$ molecule is abundant,
whilst HCN may be more abundant when the 3.8 $\mu$m band is absent. This is
corroborated by the observation that the 3.57 $\mu$m band due to HCN is seen
both in objects with very strong 3.8 $\mu$m absorption and in objects without
that C$_2$H$_2$ band. The 3.57 $\mu$m shows no dependence on luminosity or
mass-loss rate, suggesting that its abundance is primarily set by the
initially available nitrogen with little evidence for additional nitrogen
produced by the CNO cycle in more massive AGB stars (those would probably not
become carbon stars anyway due to Hot Bottom Burning). The strength of the 3.8
$\mu$m band can thus be taken as a measure for the carbon overabundance
resulting from the carbon produced in the star itself, whilst the 3.57 $\mu$m
band measures the (initial) nitrogen abundance.

The 3.1 $\mu$m band sometimes shows a strong red wing of absorption extending
up to 3.5 $\mu$m. Although this could be due to the HNC isomer (Harris et al.\
2003), it is not expected to form in great quantities under chemical
equilibrium conditions. Unless stellar pulsation causes departures from
chemical equilibrium that result in enhanced HNC abundances, it is more likely
that the broad absorption is due to a high excitation temperature, closer to
the stellar surface. The excitation conditions almost certainly depend on the
strength of pulsation, and may vary as a function of phase in the pulsation
cycle. We find evidence for the excitation temperature to be on average lower
in stars with longer periods and/or larger amplitudes, indicative of a more
extended molecular atmosphere.

We have investigated in some detail the nature of the relatively sharp
absorptions at 3.70--3.78 $\mu$m. Of these, the 3.70 $\mu$m ``band'' is the
easiest to measure reliably. The 3.70 $\mu$m and associated sharp features are
seldomly as prominent in the SWS spectra of galactic carbon stars, suggesting
that they may be indicative of the C enhancement seen in LMC carbon stars. The
3.70 $\mu$m band is generally absent at low mass-loss rates whilst the 3.57
$\mu$m band is often quite strong in those stars, suggesting that the 3.70
$\mu$m absorption is not due to HCN. Both CH and C$_2$H$_2$ have features in
this spectral region. No CH absorption --- either around 3.75 $\mu$m or
elsewhere in the spectrum --- is evident in any of the IR carbon stars in the
LMC. It is seen around 3.4 $\mu$m in the optically bright carbon star
MSX\,LMC\,349\,B (Fig.\ 4) and in relatively blue carbon stars in Sgr\,D
(Matsuura et al.\ 2005), but in neither case does absorption at 3.8 $\mu$m
occur. This suggests that in optically bright carbon stars CH dominates over
C$_2$H$_2$ whilst in IR carbon stars the situation is reversed --- probably a
consequence of the higher C overabundance and/or denser atmosphere promoting
the formation of larger carbon-chains. The 3.70 $\mu$m band is strongest at
intermediate-strong 3.8 $\mu$m absorption, and these bands show an opposite
behaviour with respect to the 3.2--3.5 $\mu$m absorption. We suggest therefore
that, like the broad 3.8 $\mu$m band, the 3.70--3.78 $\mu$m absorption
components are due to C$_2$H$_2$, with excitation conditions determining their
relative strengths in the sense of the sharp features dominating at lower
excitation temperature in an extended atmosphere and the broad absorption
dominating at higher excitation temperature in lower (or shocked) layers of
the atmosphere.

\subsection{The assembly of larger molecules and dust grains}

The properties of the molecular atmosphere amongst the IR carbon stars in the
LMC are very homogeneous, and in particular show little --- if any ---
variation with luminosity or mass-loss rate. For instance, although C$_2$H$_2$
seems to be abundant amongst LMC carbon stars experiencing the superwind, once
in that final evolutionary stage we were unable to identify a strong
correlation between the C$_2$H$_2$ absorption and the mass-loss rate. The
subsequent production of PAH molecules and carbon grains may be largely
constant. Veiling of the 3--4 $\mu$m spectrum by the continuum emission of
warm dust is shown to be important, but it is difficult to accurately account
for.

Strong pulsation and intense mass loss do go hand in hand, and we presented
some evidence for an extended molecular atmosphere: low excitation
temperature, saturated molecular bands, and a possible contribution from
absorption seen against the dust envelope. Although this could be thought of
as a physical link between the action of pulsation and the dust-driven wind,
no clear cause-effect relationship is found. This might indicate that the
pulsation and atmospheric structure have already reached the threshold
required for complete condensation into grains, leading to a ``saturation'' of
the mass-loss mechanism. That could explain the metallicity-insensitivity of
the maximum mass-loss rate attained by AGB stars (van Loon 2000, 2002).

Further insight into the molecule and dust formation should come from a more
extensive comparison between oxygen-rich AGB stars and carbon stars in the LMC
with those in the SMC, and between the properties of the molecular atmosphere
seen at optical and infrared wavelengths with the properties of the dust
grains seen at mid-IR wavelengths (Cohen 1984).

\section{Summary of conclusions}

We presented 2.9--4.1 $\mu$m spectra, obtained with ISAAC at the ESO/VLT, of a
sample of 30 optically invisible stars selected from the MSX-2MASS compilation
of Wood \& Cohen (2001). At least 28 (over 90 per cent) turn out to be carbon
stars. This confirms earlier indications (van Loon et al.\ 1997) for an
extensive population of IR carbon stars experiencing the superwind stage at
the end of the thermal-pulsing AGB. This population is now estimated to
account for about 10 per cent of the thermal-pulsing AGB lifetime of LMC
carbon stars, in agreement with findings from cluster IR objects (van Loon et
al.\ 2005b).

We presented an analysis of the molecular band strengths in the spectra of
these and several other LMC carbon stars, and conclude that in the most
evolved objects the C$_2$H$_2$ abundance is large compared to that of HCN and
CH. We interpret this as evidence for a low nitrogen abundance and a high
carbon abundance as a result of the low initial metallicity and efficient
production of carbon in LMC carbon stars, confirming and extending earlier
such evidence (van Loon et al.\ 1999; Matsuura et al.\ 2002, 2005).

No strong relationships are found between the strength of the molecular bands
and either luminosity or mass-loss rate, but we found hints for the molecular
atmosphere to be more extended at the highest mass-loss rates and strongest
pulsation, characterised by a low excitation temperature and high C$_2$H$_2$
abundance. The pulsation, molecular atmosphere and dust wind of these IR
carbon stars may have reached their maximum potential for driving mass loss.

%==============================================================================
\begin{acknowledgements}
We would like to thank the Paranal staff for excellent support at the
telescope, the anonymous referee for comments which improved the presentation
of our results, and Joana Oliveira for help and advice. JRM is supported by a
PPARC studentship and MM is supported by a PPARC fellowship. MC thanks NASA
for supporting his participation in this work through LTSA grant NAG5-7936
with Berkeley.
\end{acknowledgements}

\end{document}